\documentclass{aa}

\makeatletter

\renewcommand*\aa@pageof{, page \thepage{} of \pageref*{LastPage}}

\usepackage{natbib}
\bibpunct{(}{)}{;}{a}{}{,}%
\def\bibfont{\aa@bibliographyfont}%
\providecommand{\@LN}[2]{}

\makeatother

\usepackage{graphicx}
\usepackage{newtxtext}
\usepackage[smallerops,cmbraces]{newtxmath}
\usepackage{xcolor}
\usepackage{float}
\definecolor{xlinkcolor}{cmyk}{1,1,0,0}
\usepackage[
  bookmarks=true,         
  pdfnewwindow=true,      
  colorlinks=true,        
  linkcolor=xlinkcolor,   
  citecolor=xlinkcolor,   
  filecolor=xlinkcolor,   
  urlcolor=xlinkcolor,    
  final=true,
]{hyperref}

\usepackage{bm} 
\usepackage{textgreek} 

\usepackage[capitalize]{cleveref} 
\crefname{section}{Sect.}{Sects.}
\creflabelformat{equation}{#2#1#3} 
\crefname{enumi}{item}{items} 

\usepackage{siunitx} 
\DeclareSIUnit[number-unit-product = ]\percent{\char`\%} 

\usepackage{csquotes} 
\definecolor{blackberry}{HTML}{8D1D75}

\newcommand*{\code}[1]{\texttt{#1}} 

\newcommand*{\LCDM}{\textLambda{}CDM}

\DeclareSIUnit\parsec{pc}
\DeclareSIUnit\dex{dex}
\DeclareSIUnit\h{\mathnormal{h}}
\DeclareSIUnit\year{yr}
\DeclareSIUnit\years{yrs}
\DeclareSIUnit\arcsec{arcsec}
\DeclareSIUnit\arcmin{arcmin}
\DeclareSIUnit\Msun{M_\odot}
\DeclareSIUnit\Rsun{R_\odot}
\DeclareSIUnit\Lsun{L_\odot}
\DeclareSIUnit\Rvir{\mathnormal{R}_\mathrm{vir}}
\DeclareSIUnit\Rhalf{\mathnormal{R}_{1/2}}
\DeclareSIUnit\erg{erg}
\DeclareSIUnit\angstrom{\text{Å}}

\newcommand*{\Msun}{\ensuremath{\mathrm{M}_\odot}} 
\newcommand*{\Rsun}{\ensuremath{\mathrm{R}_\odot}} 
\newcommand*{\Lsun}{\ensuremath{\mathrm{L}_\odot}} 
\newcommand*{\rhalf}{\ensuremath{r_{1/2}}}
\newcommand*{\rvir}{\ensuremath{r_\mathrm{vir}}} 
\newcommand*{\Rhalf}{\ensuremath{R_{1/2}}} 

\begin{document}

\title{Living the stream — Properties and progenitors of tidal shells and streams around galaxies from Magneticum}
\titlerunning{Properties and progenitors of tidal shells and streams} 

\author{
    Johannes Stoiber\inst{\ref{inst:usm},\ref{inst:origins}}\thanks{jstoiber@usm.lmu.de}
    \and
    Lucas M.\ Valenzuela\inst{\ref{inst:usm}}
    \and
    Rhea-Silvia Remus\inst{\ref{inst:usm}}
    \and
    Lucas C.\ Kimmig\inst{\ref{inst:usm}}
    \and
    Jan-Niklas Pippert \inst{\ref{inst:mpe}}
    \and 
    Elisabeth Sola\inst{\ref{inst:cambridge}}
    \and 
    Klaus Dolag\inst{\ref{inst:usm},\ref{inst:mpa}}
}
\authorrunning{J\ Stoiber et al.}

\institute{
    Universitäts-Sternwarte, Fakultät für Physik, Ludwig-Maximilians-Universität München, Scheinerstr.\ 1, 81679 München, Germany\label{inst:usm}\\
    \email{jstoiber@usm.lmu.de}
    \and
    Excellence Cluster ORIGINS, Boltzmannstraße 2, 85748 Garching, Germany\label{inst:origins}
    \and 
    Max-Planck-Institut für Extraterrestrische Physik, Gießenbachstr.\ 1, 85748 Garching, Germany\label{inst:mpe}
    \and 
    Institute of Astronomy, Madingley Rd, Cambridge CB3 0HA, UK\label{inst:cambridge}
    \and
    Max-Planck-Institut für Astrophysik, Karl-Scharzschild-Str.\ 1, 85748 Garching, Germany\label{inst:mpa}
}

\date{Received XX Month, 20XX / Accepted XX Month, 20XX}

\abstract
{Stellar shells and streams are remnants of disrupted satellite galaxies visible around host galaxies. Advances in low-surface-brightness observations and increasing resolution of cosmological simulations now allow investigating the properties and origin of these features.
}
{The stellar population properties of shells and streams, that is, metallicity, age, and velocity dispersion, are investigated to infer the progenitor galaxies properties from such features.}
{We employed the hydrodynamical cosmological simulations Magneticum Pathfinder to extract these properties and identify the progenitors of the shells and streams. We compared to observational results from surveys and individual galaxies, matching and testing the methodology used in observations.}
{
Mock observations of shells and streams agree well with observational data regarding their morphology and spatial distribution. We find that both types of features are associated with localized depressions in stellar velocity dispersion compared to the surrounding regions. They are not as clearly distinct in metallicity and ages, though overall shells and more metal rich and streams are younger. We confirm results from idealized models that shells form more commonly from radial major mergers but also through minor mergers, while streams usually form from minor mergers on more circular orbits. We do not find the widths of streams to correlate with the half-mass radii of their progenitors, but the progenitors follow the mass-metallicity relation, which justifies mass estimates of progenitors based on this relation. On average, the masses measured for shells and streams approximately corresponds to 20\% of the progenitor mass. Finally, we introduce a class of very young, star-forming streams, which originate from in-situ star formation rather than from disruption of satellite galaxies.
}
{Measuring stellar population properties of shells and streams provides the means to reconstruct the progenitor properties, and especially distinguish those streams that are not made through the disruption of a progenitor galaxy but formed in-situ from a large-scale gas reservoir.}


\maketitle
%

\section{Introduction}
\label{sec:introduction}

Galaxy evolution is driven by mergers, where the properties of the resulting galaxy are strongly dependent on the merger mass ratio \citep[e.g.][]{toomre&toomre72, hoffman+10, moody+14, karademir+19}. These events are typically subdivided into major mergers, minor mergers, and mini mergers \citep[e.g.][]{schulze+20}. The remains of such a merger can be found in the low-surface-brightness (LSB) outskirts of a galaxy and, for example, appear as tidal shells, streams, or tails \citep[e.g.][]{arp66, toomre&toomre72, quinn84, amorisco15, mancillas+19, veracasanova+22}. Additional tracers of a galaxy's formation history are stellar age and metallicity, as well as their spatial distribution \citep[e.g.][]{sextl+24, venturi+24}. A combined analysis of tidal features and their stellar population properties, such as their ages and metallicity, is crucial in understanding galaxy evolution \citep[e.g.][]{fensch+2020}.

The number of observation of LSB tidal features has vastly increased over the past decade \citep[e.g.][]{duc+15, bilek+20, sola+22, rutherford+24, martinez_delgado+23a, martinez_delgado+25, sola+25, sola+25b}. An important statistic in their study is their incidence, which tells us how many galaxies experienced strong tidal effects during the last few Gyr of their evolution. Recent estimates are found to be up to ${\sim}30\%$ \citep{bilek+20, sola+22} at a surface brightness limit of $29 \mathrm{\ mag\ arcsec}^{-2}$ around early type galaxies (ETGs) in galaxy groups. This agrees well with an estimate from multiple cosmological simulations, where the mean tidal feature fraction is $35 \%$ at $30.3\mathrm{\ mag\ arcsec}^{-2}$ \citep{khalid+24}. A deeper surface brightness limit ($32\mathrm{\ mag\ arcsec}^{-2}$) is needed to reach a detection frequency of tidal features up to ${\sim}70\%$ \citep{miro_carretero+24}. In such deep observations, more than ${\sim}80\%$ of the total flux from tidal features should be detected \citep{martin+22}. 

From these vast catalogs of LSB features, more different morphologies have been identified, which can help to understand their origin. They appear in many shapes, ranging from streams, tails, shells, asymmetric halos, double nuclei \citep[e.g.][]{khalid+24}, and umbrellas \citep[e.g.][]{amorisco15, martinez_delgado+23a, martinez_delgado+23b}, to stellar halos themselves \citep[e.g.][]{sola+25}. Shells are believed to form from a merger on a radial orbit, while a stream forms on a more circular orbit, both consisting of accreted material \citep[e.g.][]{quinn84, amorisco15, karademir+19}. Tails are elongated stellar structures thought to consist of the material of the host galaxy formed during a major merger \citep[e.g.][]{sola+22}. Stellar umbrellas are features displaying an elongated structure ending in a shell \citep[e.g.][]{martinez_delgado+23a}, with one of the earliest observations of such a feature being around NGC\,4651 by \citet{Zwicky+1956}. N-body models of satellite galaxy infall on fine-tuned orbits explain their formation \citep[e.g. for NGC 922,][]{wong+06, martinez_delgado+23b}. The merger ratios for each class are of crucial interest. From cosmological zoom-in simulations \citep{martig+2012}, \citet{mancillas+19} give the estimates 1.5:1 for tidal tails, 10:1 for stellar streams, and 4:1 for shells, as well as the survival times, which are longest for streams, followed by shells ($\sim 3$ Gyr) and then tails ($\sim 2$ Gyr). \citet{pop+18} find a stellar mass merger ratio of more than 1:3 (i.e., $M_\star^\mathrm{sat}/M_\star^\mathrm{host} \geq 1/3$) for more than 50\% of shell forming mergers from the Illustris simulation \citep{Vogelsberger+14}. 

Apart from the merger ratio, many other aspects of stellar shells have been studied since their appearance around the galaxy NGC 1316 was first connected to the infall of a smaller galaxy by \citet{Schweizer80}. Further investigating this formation scenario with idealized simulations, \citet{quinn84} confirmed the consistency of stellar shells with structures formed by the radial collision of an elliptical galaxy and a disk galaxy, while non-radial encounters lead to streams rather than shells. The formation history of the prominent shell galaxy NGC 474 \citep{duc+13} has been studied in detail. \citet{bilek+22} provided a list of visible shells, and \citet{fensch+2020} estimated stellar age and metallicity of the most spectacular one from full spectral fitting of data measured by the Multi Unit Spectroscopic Explorer (MUSE) mounted at the VLT in combination with the associated globular clusters and planetary nebulae. The shell progenitor's mass was derived via the mass-metallicity relation. 

Focusing on metallicity, \citet{pop+17} found shells in the cosmological simulation Illustris \citep{Vogelsberger+14} to be more metal-rich than their surroundings and that shells from different progenitors, as well as different generation of shells, have differences in their metallicity. \citet{pop+18} conclude that these progenitors are accreted on low angular momentum orbits, which is a strict requirement for small companions, whereas massive satellites can cause shells for a larger range of impact parameters due to dynamical friction. They also found that the incidence of shells is larger for massive galaxies\citep{pop+18}. 

Similarly, stellar streams have since been investigated by analytical models \citep{amorisco15}, high resolution N-body simulations \citep{errani+15, karademir+19}, cosmological simulations \citep{mancillas+19, panithanpaisal+21, veracasanova+22, valenzuela&remus24}, and observations \citep[e.g.][]{sola+22, martinez_delgado+23a, martinez_delgado+23b, pippert+25, sola+25}. The width of a stream is more meaningful than for a shell, as it might still be related to the size of its progenitor. Recently, \citet{pippert+25} developed a tool to fit the light profile perpendicular to the elongation direction and infer the widths of streams from photometry, providing a catalog of the widths and several other properties for 15 streams and tails from the WWFI (Wendelstein Wide-Field Imager) cluster survey \citep{kluge+2020}. \citet{pippert+25} assumed a constant or increasing width or effective radius of streams when discussing possible progenitors, although they suspected the width to depend on multiple variables such as impact parameter, relative velocity, size, or mass. So far, this regime of questions has only been approached by analytical models and isolated simulations. \citet{amorisco15} determined that the widths of a stream in the orbital plane are primarily an expression of random motion escape, agreeing with \citet{errani+15}, who additionally focused on the width's dependence on the distance from the host galaxy's center and the number of pericenter passages. In their isolated N-body simulations, the widths increase with galactocentric radius, but at the same distance decrease with the increasing number of pericenter passages. \citet{erkal+16} developed a analytical model which includes the streams widths and tested it against isolated N-Body simulations. Their model predicts the width's increase with galactocentric radius. \citet{erkal+16} also showed that a stream is broader in a flattened potentials and that progenitors on polar orbits produce wider streams. 

Additionally, the age and metallicity provide further information. One of the best measured extragalactic streams is the Giant Stellar Stream at a galactocentric radius of $\sim 100$ kpc from M31, spanning a width of $\sim 25$ kpc. The stream's metallicity is higher in its center than closer to the halo stars \citep{Ibata+07, ferguson+2016}. Comparing FUV to NIR fluxes \citet{barnes+14} estimated the stellar age of the stream around M83 to be of intermediate or old age, with a stellar mass of $\sim\SI{1e8}{\Msun}$ and a width between $\SI{1.6}{\kilo\parsec}$ at the pericenter and $\SI{2.8}{\kilo\parsec}$ at the apocenter. Another example is provided by \citet{laine+16} who used stellar population synthesis to infer the age, metallicity, and mass of the stream around NGC 5907. Again, \citet{laine+16} use the mass-metallicity relation to estimate the progenitor's mass. 

Finally, the connection between tidal features and the internal kinematics of early-type galaxies was recently explored in more detail. Regarding the line-of-sight velocity, a connection between prolate rotation \citep{krajnovic+11} and the appearance of shells was found both in observations \citep{ebrova+21} and in simulations \citep{valenzuela&remus24}. More broadly \citet{bilek+22} found that tidal disturbances occur more often around slowly rotating ETGs. Specifically, shells more likely appear around slowly rotating galaxies \citep{valenzuela&remus24, yoon+24, rutherford+24}. Using the Magneticum Pathfinder simulations \citet{valenzuela&remus24} find that for $30\%$ of the shell galaxies, the shell-causing mergers also slow down the rotation. \citet{yoon+24} investigated 1244 galaxies from MANGA and deep DESI Legacy Survey images and reported that half of the ETGs that exhibit shells are slowly rotating, which further solidifies their radial merger origin. \citet{rutherford+24} find such a correlation within SAMI galaxies the rotational support and the incidence of shells only for galaxies younger than $\SI{10.8}{\giga\year}$. Regarding the velocity dispersion, \citet{valenzuela&remus24} notice for a case study that shells appear as a depression in velocity dispersion, while they do not appear in velocity maps. This raises the question whether it is possible to identify shells and streams based on their stellar population properties compared to those of their surrounding galactic halo. The velocity dispersion of stellar streams was analyzed by \citet{errani+15} in idealized simulations, who found that the velocity dispersion of a stream is higher than within its progenitor for most models and is highly dependent on the orbit, number of orbit periods, and the distance from the center of the host galaxy. It decreases with increasing galactocentric radius and increasing number of pericenter wraps.

\begin{table*}
    \centering
    \caption{Summary statistics of the identified tidal features. $N_\mathrm{galaxies}$: number of galaxies that exhibit the respective feature. $N_{\geq 1}$: number of galaxies for that a feature could be identified in at least one projection. $N_{=1}, N_{=2}, N_{=3}$: numbers of galaxies for which a feature could be identified in exactly one, two or three projections. $N_\mathrm{edge-on}$, $N_\mathrm{faceon-on}$, $N_\mathrm{side-on}$: numbers of features found in each projection (counted individually). $N_\mathrm{progenitor}$: number of progenitors found for the features visible in the face-on projection.}
    \begin{tabular}{c|c|c|ccc|ccc|c}
    & $N_\mathrm{galaxies}$ & $N_{\geq 1}$ & $N_{=1}$ & $N_{=2}$ & $N_{=3}$ & $N_\mathrm{edge-on}$ & $N_\mathrm{face-on}$  & $N_\mathrm{side-on}$ & $N_\mathrm{progenitor}$  \\
    \hline 
    Shells & 24 & 24 & 3 & 17 & 4 & 51 & 60 & 4 & 26 \\
    Streams & 66 & 56 & 11 & 29 & 16 & 51 & 57 & 40 & 40 \\
    \end{tabular}
    \label{tab: identify}
\end{table*}

As both the LSB features, as remnants of major and minor mergers, and the stellar age and metallicity are tracers of the formation history of a galaxy, we used the cosmological simulation \emph{Magneticum Pathfinder} \citep{dolag+25}, building on a study by \citet{valenzuela&remus24}, to analyze and compare the ages and metallicity of stellar shells and streams, as well as the velocity dispersion and structural properties. We study the possibility to identify shells and stream from the stellar velocity dispersion, the stellar age, and the metallicity. Additionally, we directly study the formation history of these features by identifying their progenitor galaxy. We find that shells and streams are a depression in velocity dispersion maps, but their widths are not correlated to the size of their progenitor. We present the origins of shells and streams, showing that shell progenitors are typically more massive than those of streams, leading to merger orbits that are more radial. We compared the widths of streams to their progenitors sizes to answer whether the width of a stream is larger, equal, or smaller than the size of its progenitor for the first time from cosmological simulations. We find the projected stream width to not correlate with the progenitor size. AS the mass-metallicity relation is a widespread method to estimate the stellar metallicity of a tidal feature's progenitor, we investigate the robustness of this approach, and find that shell and stream progenitors follow the general mass-metallicity relation. In addition, we identify a new class of young streams that form in-situ from a pre-existing gaseous ring around the host, likely through an instability triggered by a close encounter.

\section{The Magneticum Pathfinder simulations}
\label{sec:Magneticum}

The galaxies analyzed in this work were selected from the \emph{Magneticum Pathfinder Simulations}\footnote{\url{www.magneticum.org}} \citep{dolag+25}, which is a suite of cosmological hydrodynamical simulations performed with the Tree/SPH code \textsc{Gadget-3}, a successor of \textsc{Gadget-2} \citep{springel+01a}. It includes various improvements, for example, the used kernels \citep{dehnen&aly12, beck+16}, artificial viscosity \citep{dolag+05b}, and passive magnetic fields \citep{dolag+09}. The underlying cosmology is standard \LCDM{} using parameters adopted from the seven-year results of the Wilkinson Microwave Anisotropy Probe (WMAP7,\citealp{komatsu+11}). The density parameters are $\Omega_b = 0.0451$, $\Omega_M = 0.272$, and $\Omega_\Lambda = 0.728$, for baryons, matter, and dark energy. The reduced Hubble parameter is $h = 0.704$ and the normalization of the fluctuation amplitude at \SI{8}{\mega\parsec} is $\sigma_8 = 0.809$. The included physics is described in detail by \citet{dolag+25}.

We used Box4 (uhr), which has a side length of $\SI{48}{\mega\parsec\per\h}$. The masses of particles are $m_\mathrm{dm} = \SI{3.6e7}{\Msun\per\h}$, $m_\mathrm{gas} = \SI{7.3e6}{\Msun\per\h}$, and on average $\langle m_\star \rangle = \SI{1.4e6}{\Msun\per\h}$, for dark matter, gas, and stars. The softening lengths are $\epsilon_\mathrm{dm} = \epsilon_\mathrm{gas} = \SI{1.4}{\kilo\parsec\per\h}$, and $\epsilon_\star = \SI{0.7}{\kilo\parsec\per\h}$. The galaxies in this box have been found to match well with observations regarding angular momentum \citep{teklu+15}, kinematics \citep{schulze+18, van_de_sande+19, schulze+20}, dynamics \citep{remus+17, teklu+17, harris+20}, and the in-situ components' fractions \citep{remus&forbes22}. The stellar metallicity gradients have been studied, compared, and were found to match well with those of globular clusters in ETGs \citep{forbes&remus18, kudritzki+21}. 

Main halos and subhalos were identified by \textsc{subfind} \citep{springel+01b}, which applies a friends-of-friends (\textsc{fof}) algorithm \citep{davis+85} and is modified to include the baryonic component \citep{dolag+09}. 

Merger trees were built using \textsc{L-BaseTree} \citep{springel+05} to identify the subhalos at higher redshifts that end up forming a specific subhalo at $z = 0$. The most massive one among these at a given snapshot is called the first progenitor at the snapshot's redshift.  

\begin{figure*}
    \centering
    \includegraphics[width = \textwidth]{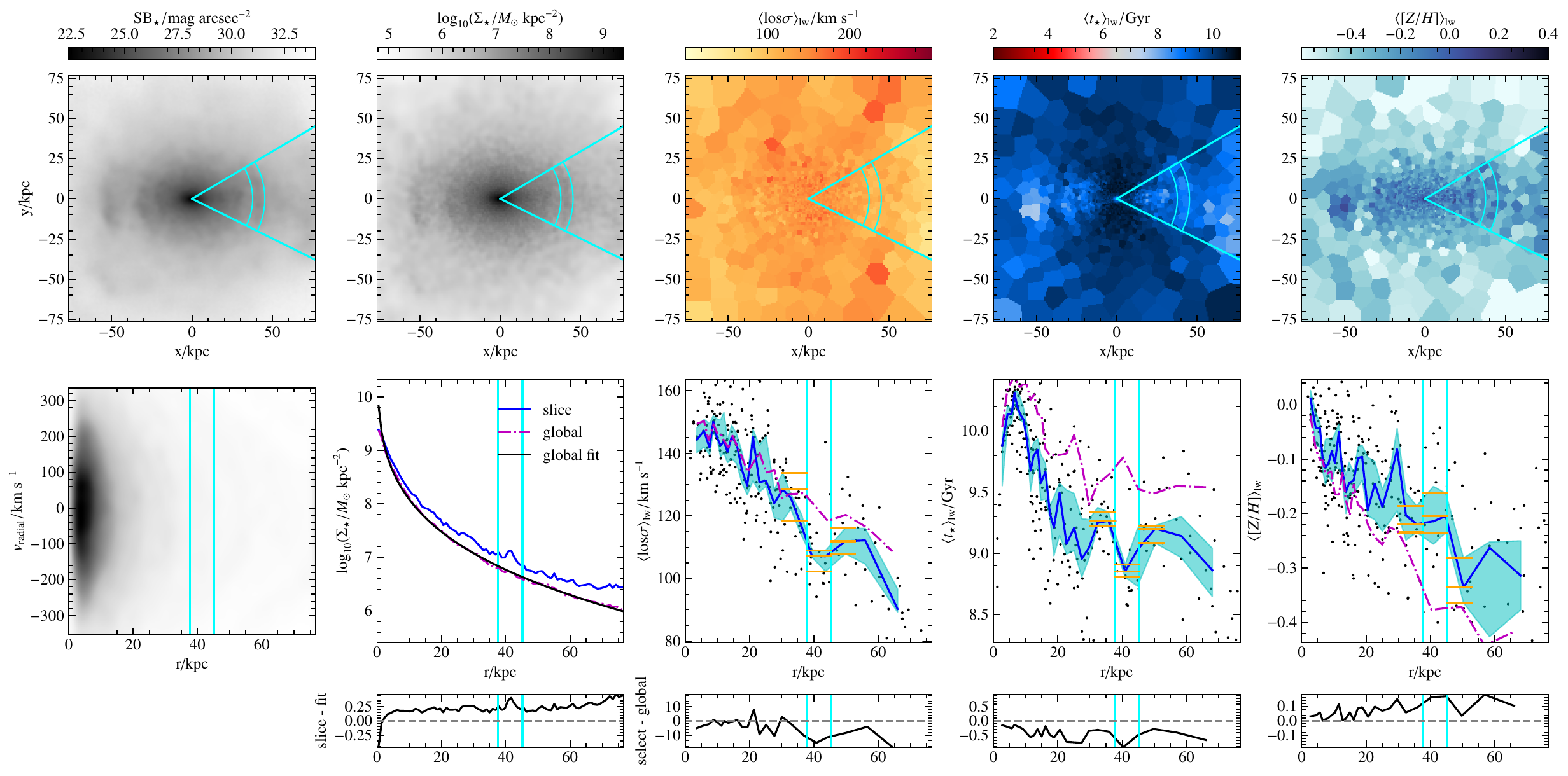}
    \caption{Top row: Stellar mock surface brightness map, 2D-binned stellar surface density map, luminosity-weighted Voronoi-binned maps of velocity dispersion $\langle \mathrm{los}\sigma \rangle_\mathrm{lw}$, stellar age $\langle t_\star \rangle_\mathrm{lw}$ and stellar metallicity $\langle [Z/H] \rangle_\mathrm{lw}$. All are in the face-on projection and have an extent of $14\rhalf \times 14 \rhalf$. Here $r_{1/2}$ is the stellar half-mass radius. Overplotted are the contours of the slice and the radial range of the shell as cyan lines, and the shape ellipses at $1\rhalf$ and $3\rhalf$ in blue.
    Middle row: Radial velocity-radius phase space distribution of stellar particles colored by a Gaussian kernel density estimation, stellar surface density profile within the slice (blue solid line), for the global galaxy (magenta dash-dotted line), and a fit of a Sérsic profile to the global profile (black solid line). The remaining three panels show the radial profiles of the velocity dispersion $\mathrm{los}\sigma$, stellar age $\langle t_\star \rangle_\mathrm{lw}$, and metallicity $\langle [Z/H] \rangle_\mathrm{lw}$. In each panel, the black dots represent individual Voronoi-bins, and the blue line represents median values within radial bins, which each contain 50 Voronoi-bins, within the slice. The cyan contours stretch from the 0.32 to 0.67-quantile within these radial bins. The magenta dash-dotted line represents the global radial profile, calculated in the radial bins defined by the slice profile. The orange horizontal lines depict the extremum (minimum for $\langle \mathrm{los}\sigma \rangle$ and age, maximum for metallicity) of the median radial profile (blue line) within the radial range of the shell (cyan vertical lines) and the opposite extremum (maximum for $\langle \mathrm{los}\sigma \rangle$ and age, minimum for metallicity) within the radial ranges of width $w_\mathrm{shell}$ adjacent to the shell, together with their 0.32 and 0.67 quantiles. Bottom row: The difference between the profile of each property within the slice and the global profile (or a fit to it in the case of the surface density) is shown. (UID 13500)}
    \label{fig: a1-dia_shell}
\end{figure*}

\section{Identification and stellar population maps}
\label{sec:ident}
\label{subsec:mocks&maps}

\begin{figure*}[htbp]
    \centering
    \includegraphics[width = \textwidth]{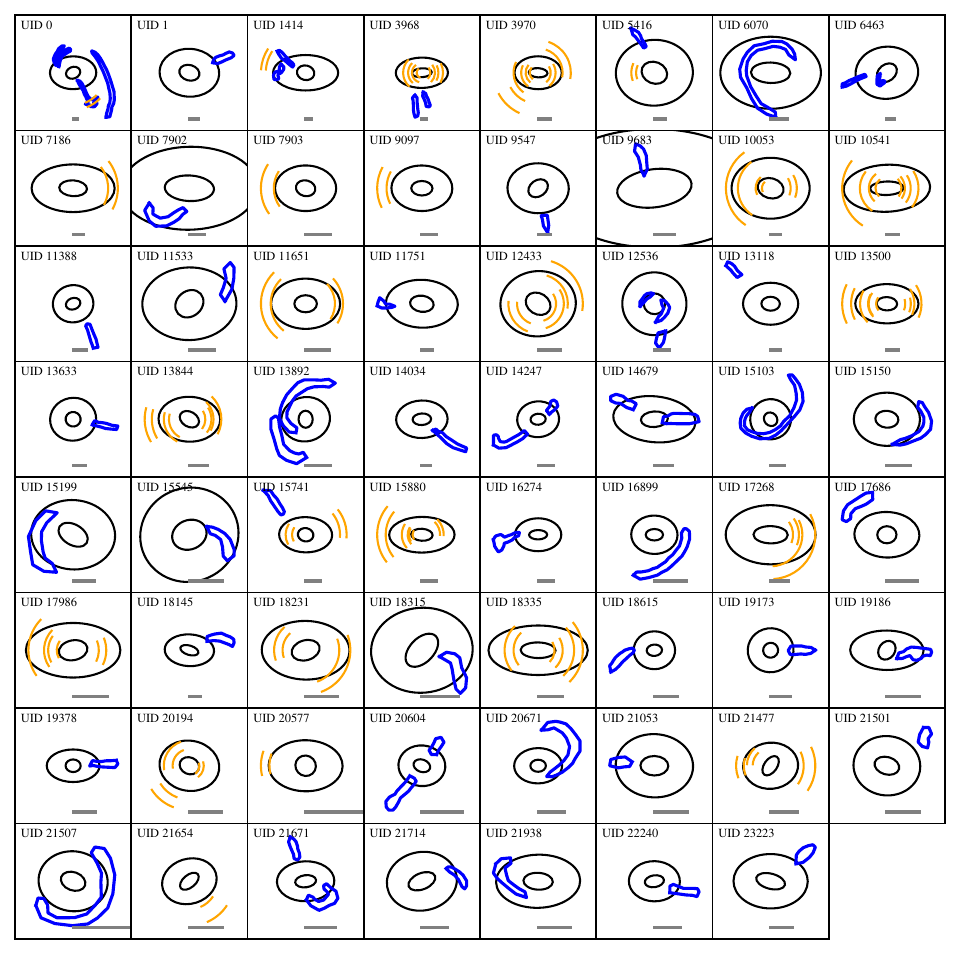}
    \caption{Overview of the shapes of shells (orange) and streams (blue). The black ellipses in the background represent the host galaxy's shapes at $1\rhalf$ and $3\rhalf$. The gray horizontal line is a scale bar of $\SI{15}{\kilo\parsec}$. All galaxies are rotated face-on. Inspired by \citealp{sola+22} (their Fig.\,7).}
    \label{fig: 1-all_shapes}
\end{figure*}

The initial sample of galaxies exhibiting shells and streams was identified by \citet{valenzuela&remus24}. Using a three-dimensional visual inspection, they found 24 galaxies with shells and 66 galaxies with streams above $\log_{10} (M_\star / \Msun) \geq 10$ at $z = 0.07$. As this classification was done solely visually no merger ratios were considered for the classification, but were calculated in this work (see \cref{sec:ratio-gas-morphology}). Most galaxies are ETGs, with only a few stream galaxies being LTGs. A considerable number of stream galaxies also belongs to an intermediate class, which includes S0 galaxies \citep{teklu+15, valenzuela&remus24}. We centered and rotated the galaxies to align their principle axes with the coordinate system according to the procedure presented by \citet{valenzuela+24}. We created stellar mock observations (r-band), using the method presented by \citet{martin+22}, for each galaxy and projection (edge-on, face-on, side-on), because not all features are visible in all projections \citep{pop+18, mancillas+19}. The limiting surface brightness and the spatial resolution were chosen to be $ \mu_\mathrm{lim} = 30.3 \mathrm{\ mag\ arcsec}^{-2}$ and $0.2 \mathrm{\ arcsec\ pixel}^{-2}$, which mimics the Legacy Survey of Space and Time (LSST) \citep{martin+22, khalid+24}. Most LSB features are typically found in the outskirts of their host galaxies, but dynamically young shells are still close to the center \citep{bilek+15}. Therefore, we used two different contrast levels to better identify features at different radii following the procedure by \citet{pop+18}. At a high contrast, the features in the outskirts are visible, while at a low contrast, the center is not obscured. 

\cref{tab: identify} summarizes the statistics of features found using these three different projections. For all shell galaxies, a shell was found in at least one projection. In agreement with \citet{pop+18} for the Illustris simulations, most shells are identified in two projections ($\sim 71\%$). But for only four ($\sim 17\%$) shell galaxies, shells were identified in all projections, whereas \citet{pop+18} found that to be the case for $\sim 40\%$ of their sample. This is due to the different choices of projections (random vs. edge-on, face-on, side-on). For three ($\sim 13\%$) shell galaxies, shells were visible in only one projection. We find that shells are mostly identified in the edge-on (51 individual shells) and face-on (60 individual shells) projections  but are barely visible in the side-on (4 individual shells) projection. Generally, a shell does not cover an entire sphere, but only a spherical cap propagating outward parallel to the major axis of the host galaxy. In the line of sight parallel to the major axis, one only sees the top of the cap, while possible remaining shells are hidden behind it. Because of such projection effects, one also has to be careful to not mistake singular shells, viewed orthogonally to their propagation, as streams. 

The same trend is true for streams, where most are identified in two projections ($\sim$ 52\%), second most in all projections ($\sim$ 29\%), and 11 ($\sim 20\%$) in only one projection. For 10 galaxies classified as stream galaxies, the stream could not be found in any of the three projections. Here, the difference in the numbers of identified features between edge-on (51 individual streams), face-on (57 individual streams) and side-on (40 individual streams) is not as extreme as for shells because they do not affect the shape of the host galaxy as much. Also \citet{mancillas+19} reported that the dependence on projection is much more pronounced for shells compared to streams. For the remaining analysis, the face-on projection was used because most features are identified in this projection. 

We analyzed several properties of shells and streams. \citet{valenzuela&remus24} noted that the shells around their illustrative galaxy appear as regions of comparably low velocity dispersion. We statistically analyzed and extended this finding to all shells and streams. The stellar ages and metallicities of shells and streams were also determined. For this purpose, Voronoi-binned \citep{cappellari&copin03} maps of said properties were utilized. This is done to be more comparable to Integral Field Unit (IFU) measurements employing stellar population synthesis models to study stellar population properties, as these methods often use similar binning methods. Additionally, we used stellar surface density maps, the aforementioned stellar mock observations, and the radial velocity-radius phase space, to aid the detection. 

We measured the properties of shells in radially binned profiles produced from the Voronoi-binned maps. The radial profile was calculated from the bins within a slice of the map where a shell is located. This slice was visually identified in the 2D projected stellar mock observations. Due to their more complicated morphology and position within the host galaxy, which does not necessarily have a clear reference to the center of the host galaxy, no radial profiles were produced for stellar streams. Instead, the 2D mock observations were used to select the stream within a polygon. Their properties were calculated as the median of the Voronoi-bins within the polygon. If not stated otherwise, all properties are luminosity-weighted as indicated by the subscript $\langle \cdot \rangle_\mathrm{lw}$. 

\begin{figure*}
    \centering
    \includegraphics[width = \textwidth]{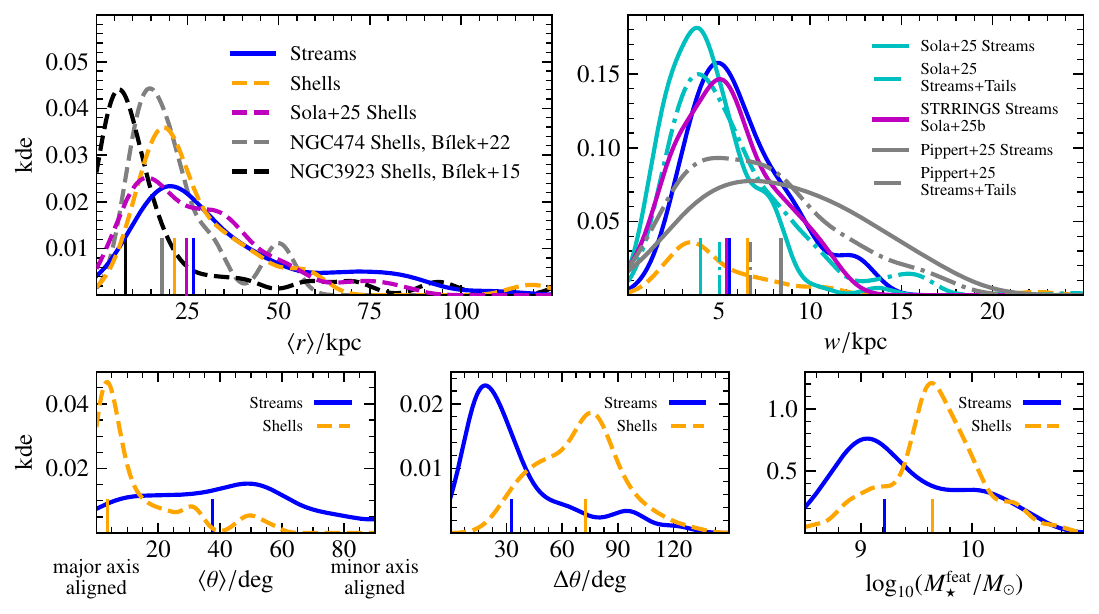}
    \caption{Top row: Distribution of the mean radii $\langle r \rangle$ and widths $w$ of shells (orange dashed lines) and streams (blue solid lines). The magenta distribution in the left panel represents shell radii measured by \citet{sola+25}. Also included are the shell radii of NGC\,474 (gray) as measured by \citet{bilek+22} and shell radii of NGC\,3923 (black) as measured by \citet{bilek+15}. In the right panel, the widths (annotated) of a sample of streams (cyan solid line) and a sample of streams combined with tails (cyan dash-dotted line) measured by \citet{sola+25} is shown, as well as measurements of the widths (Gaussian fit) from residual images provided by the STRRINGS project \citep[magenta solid line,][]{sola+25b}. Additionally, the widths (FWHM of a Gaussian fit; described in \cref{subsec:streampy}) of a sample of 9 streams (gray solid line) and a sample of the streams combined with 4 tails (gray dash-dotted line) as measured by \citet{pippert+25}.
    Bottom row: Distributions of the mean angular position $\langle \theta \rangle$ (i.e., angle between the center of the feature and the major axis), angular extent $\langle \Delta \theta \rangle$ (i.e., opening angle of the shell-defining slice or the maximal pairwise angular separation of the points of the stream enclosing polygon), and the projected stellar mass $M_\star$ of shells and streams. The short vertical lines represent the respective medians of each distribution. A Gaussian kernel was used to estimate the density of each distribution, using the KernelDensity.jl packages from the \url{https://juliastats.org} project.} 
    \label{fig: 2-stats}
\end{figure*}

Figure \ref{fig: a1-dia_shell} illustrates the method used to gain the properties of a shell. A slice in which a shell lies is visually selected in the 2D face-on projection (top row, left panel). Subsequently, we chose the radial range enclosing the shell. This radial range stretches from a radius $r_\mathrm{in}$ to a radius $r_\mathrm{out}$, from which its widths $w_\mathrm{shell} = r_\mathrm{out} - r_\mathrm{in}$ is computed. All particles within this slice and radial range are used to calculate the projected mass of the shell. Within the slice, the Voronoi bins were grouped together to compute a median radial profile (blue line in the middle row, third to fifth panels) where the $1-\sigma$ scatter is indicated by the cyan area. The minimum (for velocity dispersion and age) or maximum (for metallicity) of this profile in the radial range of the shell is chosen as the velocity dispersion, age, or metallicity of the shell (orange horizontal lines). The approach of choosing different extrema is justified by our prior visual inspection of all profiles and our expectation based on findings of low velocity dispersion within shells by \citet{valenzuela&remus24} and high metallicity within shells by \citet{pop+17}. The uncertainty of this value is given by the 0.32- and 0.67-quantile at the same position (orange horizontal lines). A measure of the properties of the surroundings of the shell is computed in the radial ranges $r_\mathrm{in}-w_\mathrm{shell}$ to $r_\mathrm{in}$ and $r_\mathrm{out}$ to $r_\mathrm{out} + w_\mathrm{shell}$, where $w_\mathrm{shell} = r_\mathrm{out} - r_\mathrm{in}$. We choose this characterization because we want to compare the feature to its local stellar halo, not the whole galaxy, which includes the center. There, the opposite extremum and the respective quantiles were chosen (orange horizontal lines) to estimate the difference between the features and their surroundings. Additionally, a surface density profile (bottom row, second panel) was computed within the slice (blue line) and for the whole galaxy (magenta dash-dotted line). We fitted a Sérsic profile $\log_{10}\Sigma(r) = \log_{10}\Sigma_0 - k\times r^{1/n_\mathrm{Sersic}}$ (black line) to the global profile. The difference of the profiles within the slice and the fitted profile is used to identify the shell as an excess in this difference (bottom row, left panel). The illustrative galaxy shown in \cref{fig: a1-dia_shell} (UID 13500) exhibits a stellar shell at $r \sim \SI{42}{\kilo\parsec}$, which is clearly younger and more metal-rich than its surroundings and has a lower velocity dispersion. This shell was chosen arbitrarily, as this galaxy has additional shells (see \cref{fig: 1-all_shapes}, \cref{app:second_appendix}).

Due to their more complicated morphology and position within the host galaxy, which has no clear reference to the center of the host galaxy, no radial profiles were produced for stellar streams. Instead, they were visually selected by overlaying a polygon on the stellar mock observations. The surroundings are characterized as the annulus around the host galaxy which inner and out radius is the innermost and outmost point of the stream. The widths of a stream $w_\mathrm{stream}$ is visually inferred s visually
inferred by selecting two suitable vertices of the polygon. For more details, refer to \cref{app:first_appendix}. 

\section{Properties of shells and streams}
\label{sec:results1}

\subsection{Positions and shapes} 
\label{subsec:pos&shapes} 

Figure \ref{fig: 1-all_shapes} presents an overview of the positions and shapes of the identified shells and streams. The shells (orange) are mostly located along the major axis of the host galaxies (black), while streams (blue) appear in a large variety all around the galaxy. Some galaxies exhibit multiple streams (e.g., UID 13892, UID 14247, UID 20604) that appear to be part of the same structure, in other words, redrawing the course of a first stream and continuing at the same curvature beyond its boundaries would end up in the other streams.

Quantitatively, the shapes and positions are summarized in \cref{fig: 2-stats}. The distribution of the average radius $\langle r \rangle = (r_\mathrm{in} + r_\mathrm{out})/2$ (top left) of shells (orange) peaks around $\langle r \rangle \sim 20\,\mathrm{kpc}$, which is in agreement with observations from MATLAS, CFIS, VESTIGE and NGVS (magenta), where most shells appear at $ r \leq 40 \mathrm{kpc}$ \citep{sola+25}. The shells around the prominent galaxy NGC\,474 (gray) are also within this range \citep{bilek+22}, although the shells of the record holding shell galaxy NGC\,3923 (black) cover a range from the very center ($\sim \SI{2}{\kilo\parsec}$) out to over $\sim \SI{140}{\kilo\parsec}$ \citep{prieur88, sikkema+07, bilek+13, bilek+15}. The average radii of streams (blue) are slightly larger, at a median of $\langle r \rangle \sim 27 \mathrm{kpc}$. \citet{barnes+14} estimate the deprojected pericenter distance of a stream around M83 to be $r_1 = \SI{28}{\kilo\parsec}$, which lies close to the median of our sample. Using the assumption $r_1/w_1 = r_w/w_2$ ($w$ is the respective width of the stream), they estimate the apocentric distance to be $r_2 = \SI{47.8}{\kilo\parsec}$, which lies at the upper end of our sample, but is still included. The mean projected distance to a stream around NGC\,5907 is 46 kpc \citep{laine+16}, which again is at the upper end of our sample. This difference between the median shell radius and median stream radius arises because a dynamically young shell still appears close to the center \citep[e.g.,][]{bilek+22}, while streams are remnants of disrupted satellites sitting in the outskirts. We also do not make a distinction between streams that radially point towards the galaxy, where the mean radius is not a representative quantity for the position of the stream, and streams that spiral the galaxy. The maximum radius for both types of features is around $\sim 60 \mathrm{kpc}$, which might be driven by the cut-off of the displayed mock observations of 7 half-mass radii $r_{1/2}$ of the host galaxies. The median half-mass radius of the sample of all host galaxies is $\langle r_{1/2} \rangle \approx \SI{7.3\pm1.8}{\kilo\parsec}$ with $1\sigma$ uncertainties. 

\begin{figure*}
    \centering
    \includegraphics[width = \textwidth]{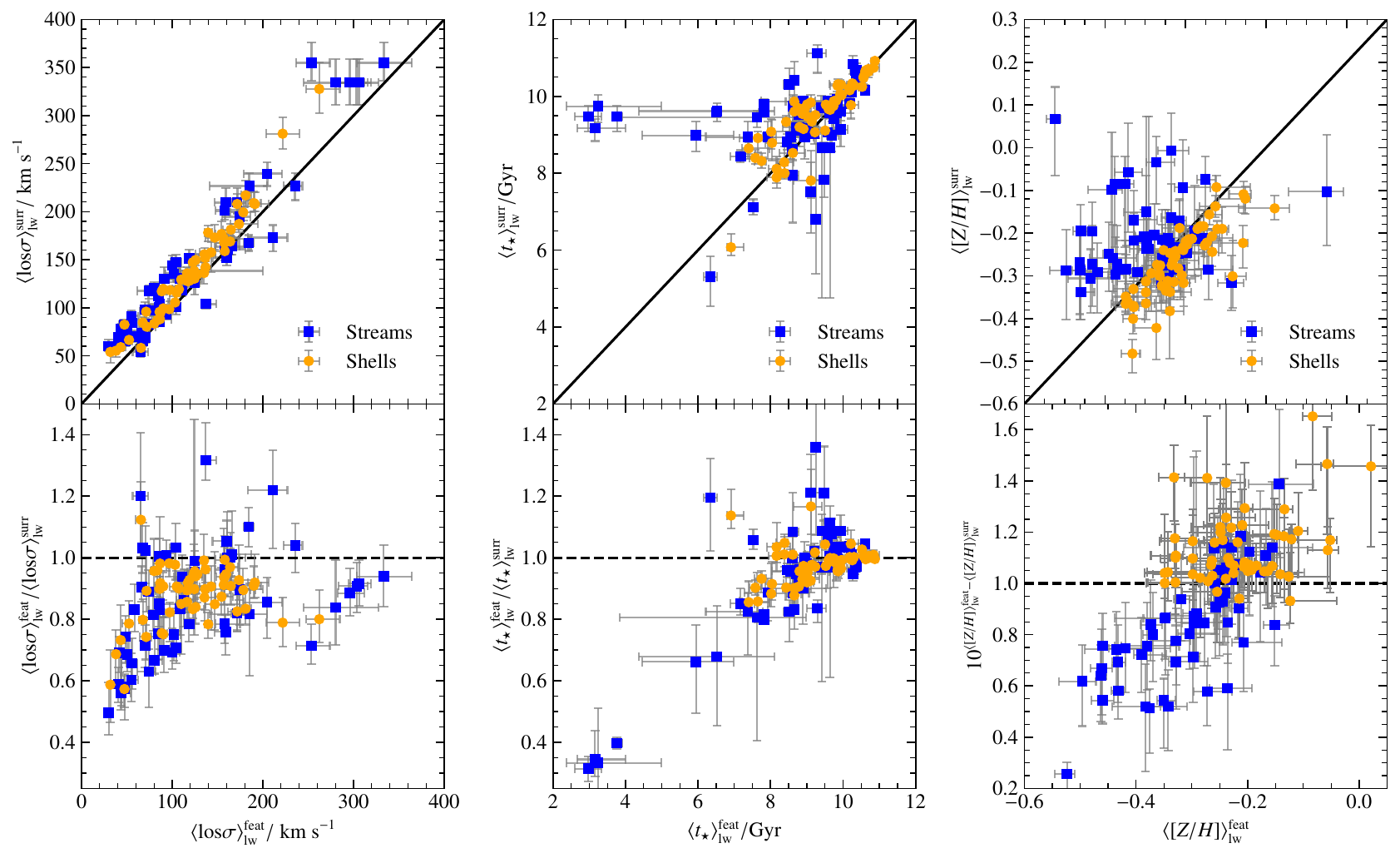}
    \caption{Top row: The respective quantity within the surroundings (sur) of shells (orange circles) and streams (blue squares) as a function of the quantity within the feature (feat). The solid black line is the one-to-one relation. From left to right: The luminosity-weighted mean stellar line-of-sight velocity dispersion $\langle \mathrm{los}\sigma \rangle_\mathrm{lw}$, the mean stellar age $\langle t_\star \rangle_\mathrm{lw}$, and the mean stellar metallicity $\langle [Z/H] \rangle_\mathrm{lw}$. Bottom row: The ratio of the respective quantity within the feature and within its surroundings as a function of the quantity within the feature. The error bars indicate the $1\sigma$ scatter within each region. }
    \label{fig: 3-props}
\end{figure*}

The width of shells ({top right of \cref{fig: 2-stats}) is on average larger than the width of streams at a median of $\langle w_\mathrm{shell} \rangle \sim \SI{6.55}{\kilo\parsec}$ and $\langle w_\mathrm{stream} \rangle \sim \SI{5.55}{\kilo\parsec}$. The stream widths are in agreement with observations of a sample of streams presented by \citet{sola+22} $\langle w_\mathrm{stream}  \rangle \sim \SI{4.1\pm2.1}{\kilo\parsec}$, a combination of streams and tails by \citet[][cyan dash-dotted line]{sola+25} $\langle w_\mathrm{stream+tail}  \rangle \sim \SI{4.0\pm2.2}{\kilo\parsec}$, and a combination of streams and tails by \citet[][gray dash-dotted line]{pippert+25} $\langle w_\mathrm{stream+tail}  \rangle \sim \SI{6.7\pm4.0}{\kilo\parsec}$. The sample of pure streams by \citet[][gray solid line]{pippert+25} is on average wider than the sample combined with tails. The simulated sample shows the best agreement with a very recent sample of streams in residual images of nearby galaxies from the DESI Legacy Imaging Surveys \citep[][STRRINGS project, magenta solid line]{sola+25b}, where the median width is $\langle w_\mathrm{stream}\rangle = 5.6^{+4.3}_{-3.3}$ kpc (uncertainties at the 0.05- and 0.95 quantile). This raises the question whether some features classified as streams are actually tails. Because our analysis is based on the classification by \citet{valenzuela&remus24}, which was done using a three dimension perspective, this is rather unlikely. While \citet{sola+22} and \citet{sola+25} use a similar method to measure the widths of streams as we do, based on visual inspection\footnote{\url{https://jafar.astro.unistra.fr/}}, \citet{pippert+25} developed a tool to fit a Gaussian profile to the stream along its entire length and calculate a median width, from the FHWM of each fitted Gaussian. A similar approach was used by \citet{sola+25b}. The annotated width is compared to a Gaussian fit width in \cref{subsec:streampy}. \citet{barnes+14} measured the width of the stream around M83 at the peri- and apocenter to be $\SI{1.6}{\kilo\parsec}$ and $\SI{2.8}{\kilo\parsec}$ using a Gaussian fit and taking the distance between the positions of $\pm20\%$ of its maximum height. These widths are at the lower end of all other considered samples.

As already suspected, the average angular position $\langle \theta \rangle$, which is the angle between the center of the feature\footnote{The angular center of a shell is the angle dividing the shell defining slice in the middle. The mean angular position of a stream is defined as $\langle \theta \rangle = (\mathrm{max}(\theta_\mathrm{polygon}) + \mathrm{min}(\theta_\mathrm{polygon}))/2$. } and the major axis, (bottom left panel of \cref{fig: 2-stats}) is concentrated close to zero for shells, while streams are uniformly distributed, showing that shells are major axis aligned while streams appear all around the host galaxy. This is likely due to a change in morphology caused by the merger that also formed the shells. The merger extended the galaxy in the infall direction, making the infall direction the major axis. Because shells are made by radial merger events \citep{karademir+19}, the infall direction is therefore also the axis along which shells appear, making the direction of the shell propagation the major axis. The prominent shells of NGC\,3923 are also oriented along the major axis \citep{prieur88}. The lack of a preferred angular position of streams is evidence for stream-forming mergers being on circular orbits that do not impact the host galaxy's shape \citep{karademir+19}. 

The angular extent $\Delta\theta$, which is the opening angle of the shell-defining slice or the maximal pairwise angular separation of the points of the stream enclosing polygon, (bottom center panel of \cref{fig: 2-stats}) of singular shells (i.e., shells on one side of the minor axis) is larger than for streams at a median of $\langle \Delta \theta \rangle_\mathrm{stream} \sim \SI{33}{\degree}$ and $\langle \Delta \theta \rangle_\mathrm{shell} \sim \SI{72}{\degree}$. Figure \ref{fig: 1-all_shapes} reveals that the smaller extent of streams is due to some streams roughly pointing towards the center of their host galaxy (e.g., UID 1), while shells always appear as a section of a circle. Streams can also be curved on themselves (e.g., UID 20671) making the angular extend smaller than a fully unraveled stream (e.g. UID 21507).

The average stellar mass (bottom right) within individual shells ($\langle M_\star \rangle = \SI{4.4e9}{\Msun}$) is larger than within individual streams ($\langle M_\star \rangle = \SI{1.6e9}{\Msun}$). This suggests that shells are more massive than streams. These are 2D projected masses, which means that the masses of all stellar particles within the respective selected areas are summed up. There likely are intruder particles that do not belong to the feature dynamically, or simply lie in the fore- or background. Estimates for observed streams are $\sim \SI{1e8}{\Msun}$ from NIR flux \citep[][for the M83 stellar stream]{barnes+14} or $\sim \SI{2.1e8}{\Msun}$ from the r-band luminosity \citep[][for the NGC\,5907 stellar stream]{laine+16}. The median stellar mass of the sample consisting of 35 streams, measured by \citet{sola+25b}, is $\SI{6.2e8}{\Msun}$ based on the g- and r-band fluxes. All compared masses of observed streams lie below our sample. They were calculated from measurements where sky background subtraction \citep{barnes+14, laine+16} or a galaxy model subtraction \citep{sola+25b} was applied. Even though careful steps have been made to not mask parts of the stream \citep{laine+16} these masses likely represent a lower bound as the faint outskirts of the streams might be subtracted as well. Due to intruder particles masses of the simulated sample are higher than the real value, while the observed masses are likely too low. Considering this, the simulated and observed value likely lie closer together. 

Overall, the radii of simulated shells and widths of simulated streams agree well with observed shells and streams. Their distributions of angular position and extent, and their stellar masses, fit the expected values. 

\begin{figure*}
    \centering
    \includegraphics[width = 0.48\textwidth]{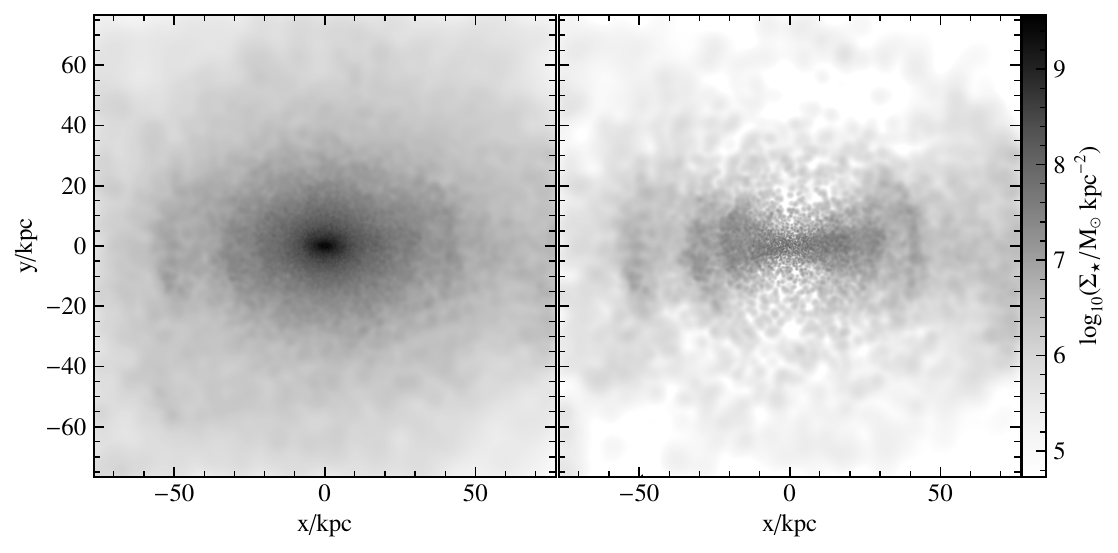}
    \includegraphics[width = 0.48\textwidth]{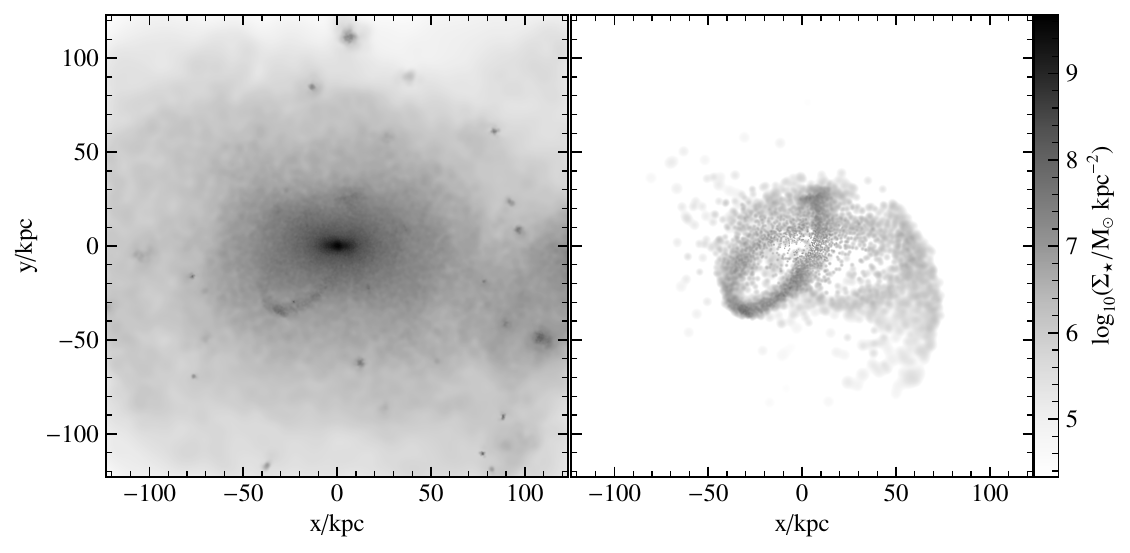}
    \caption{Leftmost panel: Stellar surface density map of an illustrative galaxy exhibiting shells (UID 13500). Second panel on the left: Stellar surface density map of the particles that used to belong to a common subhalo, that was identified to be the progenitor of the shells, traced forward to $z=0.07$. Right panels: Same as left panels but for an illustrative stream progenitor (UID 7902). }
   \label{fig: 4-decomp}
\end{figure*}

\subsection{Velocity dispersion, age, and metallicity of shells and streams}
\label{subsec:sigma-age-met}

The velocity dispersion within shells is predicted to be lower than within their surroundings \citep{valenzuela&remus24}. Furthermore, the metallicity within shells was found to be higher than within their surroundings in other simulations \citep{pop+17}. Our results regarding these parameters, together with the stellar age, are presented in \cref{fig: 3-props}. The top row presents each property (from left to right: velocity dispersion, age, and metallicity) within the surrounding galaxy as a function of each property within the feature, while the bottom row shows the ratio of both regimes. The velocity dispersion (left column) within the surroundings of shells (orange) and streams (blue) is higher than within the feature itself for most features. This is evident in the ratio, where all but one shell and 12 streams lie below unity. $68\%$ of shells and $63\%$ of streams lie further below unity than their associated $1\sigma$ scatters. This clearly shows that the velocity dispersion in both shells and streams is lower compared to the surrounding galaxy's properties, as the motion in these features is more ordered than in the galaxy itself, because a bulk of the stars still follow their progenitor's orbital path. 

The ages (central column) of shells and streams and their surroundings are, in general, similar and cluster around the one-to-one relation (black line), except for a group of eight very young streams ($\langle t_\star \rangle < \SI{7}{\giga\year}$). Only $42\%$ of shells and $46\%$ of streams are younger than their surroundings at an offset higher than their scatter. Typical ages of observed streams lie between $\SI{7}{\giga\year}$ for the Sagittarius stream in the Milky Way \citep{deboer+15}, $\SI{8.8}{\giga\year}$ for the Giant Stellar Stream around M31 \citep{brown+06}, and up to $\SI{11.9}{\giga\year}$, as measured for the stream around NGC\,5907 \citep{laine+16}. This comparison makes the young stellar streams with median ages below $\SI{7}{\giga\year}$ even more interesting. \citet{fensch+2020} measured the age of the most prominent outer shell of NGC\,474, using full spectral fitting of the shell stars and individual globular clusters associated with the shell, to be $3.55_{-0.39}^{+0.61} \mathrm{\ Gyr}$, which is younger than any shell measured in our sample, and younger than its host galaxy's (NGC\,474) core \citep[$\SI{7.65\pm1.39}{\giga\year}$,][]{mcdermid+15}, which likely is an outlier among stellar shells. 

Finally, regarding metallicity, streams are again clustered around the one-to-one relation, while shells are more metal-rich than their surroundings, consistent with \citet{pop+17}. Due to their high scatters, only $32\%$ of shells are more metal-rich than their surroundings at an offset higher than their scatter. $47\%$ of streams lie below unity in their metallicity ratios at an offset larger than their scatter. The metallicity of the Giant Stellar Stream around M31 interestingly has a high metallicity core ($\mathrm{[Fe/H]} \gtrsim -0.6$)\footnote{We compare several definitions for the term \enquote{metallicity} (e.g., [Z/H], [Z], [Fe/H], and [M/H]) assuming they are equivalent when normalized to solar values. We use the solar metallicity $(\mathrm{Z/H})_\odot = 0.0181$ measured by \citet{asplund_chemical_2009}.}, while its envelope only reaches $\mathrm{[Fe/H]} \sim -1.4$ \citep{ferguson+2016, Ibata+07}. This suggests that — at least high mass streams — are also more metal rich than their surroundings. The core metallicity of that stream lies at the lower end of our streams. The stream around NGC\,5907 has a metallicity of $\log_{10}(Z/Z_\odot) = -0.3$ \citep{laine+16}, lying right in the center of our sample. The most prominent shell around NGC\,474 has an intermediate metallicity of $\mathrm{[M/H]} = -0.83\pm0.12$, which lies significantly below our sample. On the other hand, \citet{pop+17} finds shells in the Illustris simulation that have metallicities between $\log_{10}(Z/Z_\odot) = -0.021$ and $\log_{10}(Z/Z_\odot) = 0.18$ (their target areas 2, 3, and 4), which lie significantly above our sample. The high metallicity of our simulated shells compared to their surroundings could be an imprint of the methodology, where the maximum of the binned metallicity profile within the radial range of the shell is used but the minimum within the radial range of the surroundings. This approach is justified by prior findings and our previous inspection of all radial profiles, as explained in \cref{sec:ident}, and the binning of the data gives a more robust estimate. Therefore, a physical explanation for this difference is the mass-metallicity relation: More massive galaxies, which seem to commonly be the progenitors of shells (see \cref{fig: 2-stats}), are more metal-rich. 

In summary, the velocity dispersion within streams and shells is lower than within the surrounding galaxy, confirming the results by \citet{valenzuela&remus24}, and suggesting the possibility to discover new shells and streams in highly resolved velocity dispersion maps. The ages of shells and streams are on average similar to their host galaxy's ages, but we find a sample of very young stellar streams, which will be closer examined in \cref{sec: in-situ streams}. Finally, we find that shells are more metal-rich than their surrounding, consistent with \citet{pop+17}, although our results are not as statistically significant. We did not study how this result changes if the extent of the surroundings is increased. We expect that increasing the surroundings would make this result less meaningful because the next shell might than be included in the surroundings.

\section{Progenitors of shells and streams}
\label{sec:ident-progs}

Having investigated the present features that are related to a galaxy's formation history, such as the stellar age and metallicity, it is of great importance to connect these properties to the actual formation history, which is possible within cosmological simulations.

\subsection{Identification of progenitors} 

We traced back in time the selected stellar particles within each shell and stream to identify their progenitors, and analyzed their properties. We selected all subhalos with a stellar component more massive than $\SI{1e8}{\Msun}$ ($ \gtrsim 100$ stellar particles) within a radius of $3\rvir$ around the host galaxy's first progenitors -- as identified by \textsc{subfind} -- within the 10 previous snapshots where particle data is available ($z \lesssim 0.42$, $t_\mathrm{lookback} \lesssim \SI{4.5}{\giga\year}$). The subhalo that includes the highest number of stellar particles also present in the feature at $z=0$ is selected as the possible progenitor. We inspected the stellar mass assembly history to confirm the correct identification of the progenitor. Finally, all stellar particles of the possible progenitor were traced forward, shifted into the center-of-mass frame of the host galaxy, rotated to the same viewing angle, and their distribution plotted. In \cref{fig: 4-decomp} we compare this distribution (right panel) -- as seen in the face-on projection -- to the host galaxy, where the initially selected features can be seen (left panel), for two example galaxies. This approach reveals the whole feature, including parts that have previously been hidden by the host galaxy. The results for all shell progenitors are shown in \cref{fig: a3-mosaic_shells} and for stream progenitors in \cref{fig: a4-mosaic-streams}. Using this method, 26 shell progenitors could be identified for all shells found in the face-on projection (60 in total). This means many progenitors produce multiple shells. In most cases, all shells originate from the same progenitor, but within two galaxies (UID 10541, UID 20194) there are shells produced by two different progenitors. For 43 streams out of the 57 streams found in the face-on projection, 40 progenitors were identified. Multiple galaxies exhibit more than one stream, of which the streams within three galaxies (UID 13892, 14247, 20604) are part of the same feature, while within another three galaxies (UID 1414, 14679, 21671) there are two streams produced by two different progenitors. 14 streams could not be connected to a progenitor because there was no subhalo among the previous snapshots matching a sufficient number of stellar particles. Among those are four of the eight streams younger than $\SI{7}{\giga\year}$. We will come back to these 14 peculiar cases in \cref{sec: in-situ streams}. 

\begin{figure*}
    \centering
    \includegraphics[width = \textwidth]{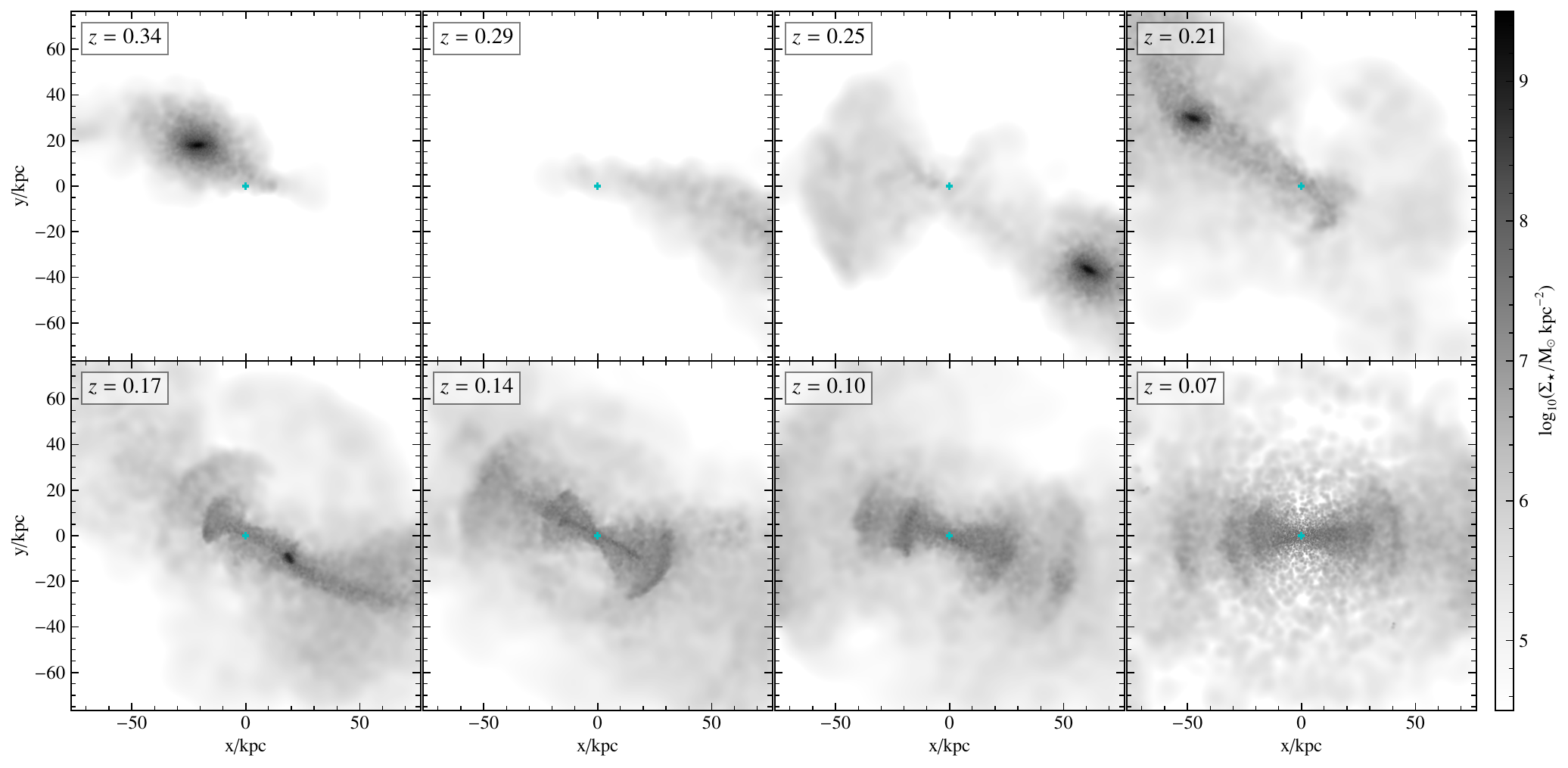}
    \caption{Evolution of an illustrative shell progenitor (UID 13500). Each panel shows the stellar surface density map of the particles belonging to the identified shell progenitor. Time evolves from the top left ($z = 0.31$) to the bottom right ($z = 0.07$). Each particle distribution is centered on the host galaxy's center (cyan plus symbol).} 
    \label{fig: 5-shell_evo}
\end{figure*}

\begin{figure*}
    \centering
    \includegraphics[width = \textwidth]{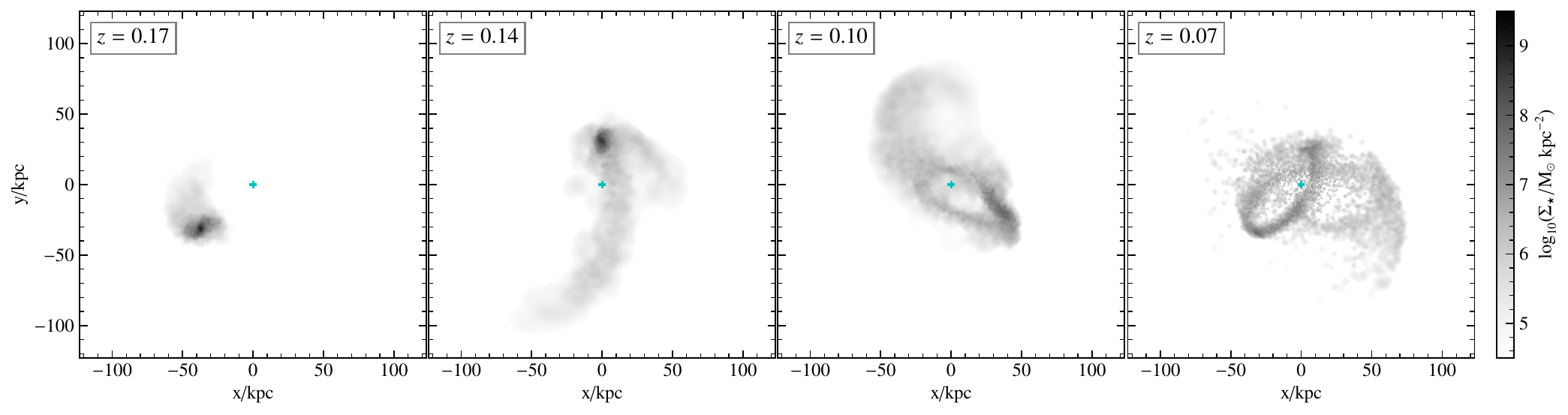}
   \caption{Same as \cref{fig: 5-shell_evo} but for an illustrative stream progenitor (UID 7902). Time evolves from the left ($z = 0.17$) to the right ($z = 0.07$). Each particle distribution is centered on the host galaxy's center (cyan plus symbol).}
    \label{fig: 6-stream_evo}
\end{figure*}

\subsection{Evolution of progenitors}

The specific identification of the progenitor as an individual subhalo allows investigating the orbit and morphology of the merger leading to the formation of shells and streams. Previously, this was done using isolated mergers \citep[e.g.][]{amorisco15, karademir+19}, using a cosmological simulation only for shells \citep{pop+18}, or by selecting specific resimulated galaxies \citep[e.g.][]{mancillas+19}. From the controlled parameter studies by \citet{amorisco15} and \citet{karademir+19} it has been shown that shells form from a radial merger while a stream progenitor falls into its host galaxy on an orbit with a high impact parameter. Figure \ref{fig: 5-shell_evo} depicts the evolution of the progenitor of a shell system from $z = 0.31$ to $z = 0.07$ in 8 snapshots, while \cref{fig: 6-stream_evo} presents the evolution of the progenitor of a stream from $z = 0.17$ to $z = 0.14$ in 4 snapshots. The center of each panel is the center of the host galaxy, marked by a cyan plus symbol. At $z = 0.31$ the shell progenitor passes very close to the center of the host galaxy and produces a first shell at $z = 0.25$. The core of the progenitor is dissolved at $z = 0.14$ ($\SI{1.73}{\giga\year}$ after the first pericenter passage), but the shell system keeps developing until $z = 0.07$. At $z = 0.17$ the stream progenitor passes the center of the host galaxy further away than the shell progenitor. Afterward, it is tidally stretched out until $z = 0.07$ ($\SI{1.18}{\giga\year}$ after the first pericenter passage). We find such pathway differences for all our identified streams and shells. This scenario is consistent with results found by \citet{amorisco15} and \citet{karademir+19} from isolated simulations, and we can thus now confirm the connection between infall orbit and feature type for a cosmological simulation context, extending on the previous results by \citet{pop+18} for shells only and the case study shown by \citet{valenzuela&remus24}. 

\begin{figure*}
    \centering
    \includegraphics[width = \textwidth]{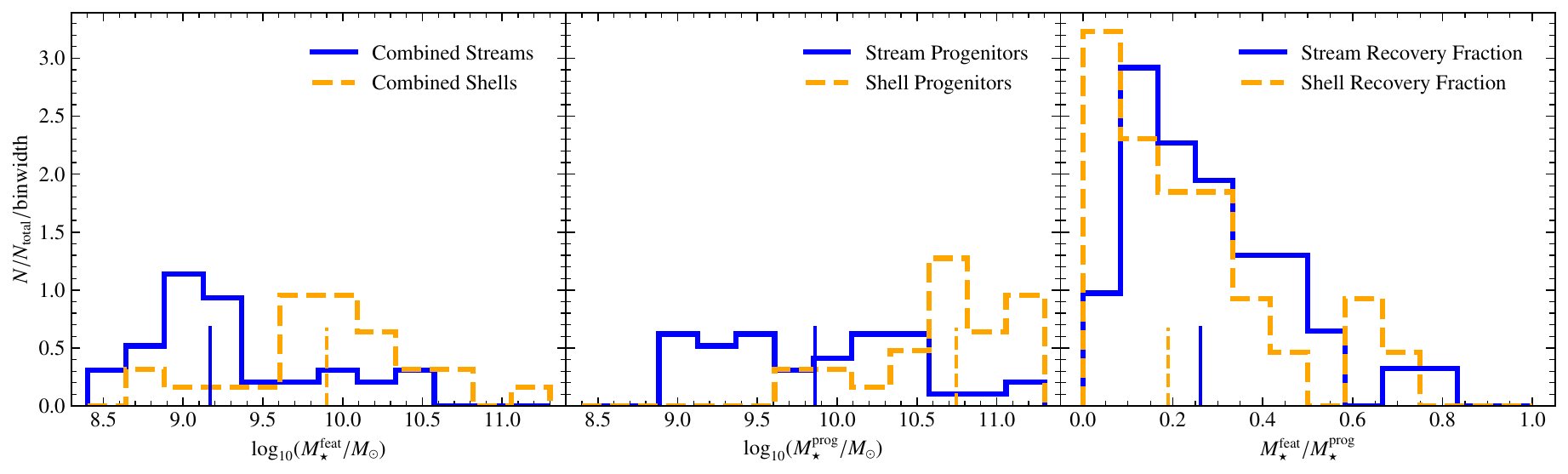} 
    \caption{Left panel: Distributions of the combined stellar masses of shells (orange dashed line) and streams (blue solid line), which were identified to originate from the same progenitor. Center panel: Distribution of the stellar mass of shell (orange dashed line) and stream (blue solid line) progenitors. Right panel: Distribution of the ratio of the stellar mass of the feature and its progenitor. The short vertical lines show the respective median. The masses of progenitors are calculated within a radius of $0.1\rvir$ at the redshift of peak stellar mass $z_\mathrm{PSM}$.}
    \label{fig: 7-masses}
\end{figure*}

\subsection{Properties of the progenitors of shells and streams} 
\label{sec:results2}

For each progenitor, several properties are determined at their redshift of peak stellar mass $z_\mathrm{PSM}$, which is the redshift at which their stellar mass is largest. At this redshift, the stellar mass of the host galaxy is also determined to calculate the merger ratio $M_\star^\mathrm{progenitor}/M_\star^\mathrm{host}$. It was confirmed that both masses stay approximately constant until the time of the merger by inspecting their stellar masses as a function of time. 

\subsubsection{Stellar masses}

Figure \ref{fig: 7-masses} shows the histogram of the summed stellar mass of all shells (orange) and streams (blue) that have the same progenitor (left panel), the stellar masses of shell and stream progenitors (middle panel), and the distribution of their ratio, the mass recovery fraction, i.e., fraction of the mass of the progenitor that is recovered by the feature (right panel). It is evident that both shells and shell progenitors are more massive than streams and stream progenitors. This trend was already present in \cref{fig: 2-stats} where individual shells are already more massive than individual streams. The combined and individual masses are contaminated by particles that belong to the host galaxy, but they are still less massive than their progenitors. The median combined shell and stream masses are $8^{+6}_{-2}\times10^9 M_\odot$ and $1^{+7}_{-5}\times10^9 M_\odot$. The median shell and stream progenitor masses are $6^{+2}_{-2}\times10^{10} M_\odot$ and $7^{+8}_{-5}\times10^{9} M_\odot$. Assuming the intruder particles are negligible, this suggests that roughly $90\%$ of the stellar mass of the progenitor is not found within the shell system or stream but is either already too diffuse or hidden by the host galaxy. The same trend is seen in the mass recovery fraction, where the maximum is located at around 10\% for both features. On average, the recovered mass fraction from the directly observed streams accounts for $26^{+10}_{-8} \%$, while for shells, the average recovery fraction accounts for $19^{+7}_{-9} \%$. 

Our median stellar stream mass is consistent with the progenitor of the M31 Giant Stellar Stream ($\SI{5e9}{\Msun}$), which was estimated by modeling the accretion of a dwarf galaxy onto M31. It is also consistent with the mass of the progenitor of the NGC 5907 stellar stream \citep[$\sim \SI{1e10}{\Msun}$,][]{laine+16}, which was estimated from the stream's metallicity and a transformation using the stellar mass-metallicity relation. Similarly, \citet{fensch+2020} find a progenitor mass of $7.4^{+14.5}_{-4.7}\times10^8 M_\odot$ for the NGC\,474 stellar shell from its stellar metallicity, which lies below our sample. They also present a second estimate where they use the i-band luminosity and assume the shells host $30\pm10$\% of the progenitor's mass: $4.1^{+11.1}_{-2.9}\times10^9 M_\odot$, which is consistent with the median shell progenitor mass of our sample. The assumption of a recovery fraction of $30\pm10$\% is reasonable, but given our results it is possible to assume a smaller fraction, which would result in a higher mass, still consistent with our shell masses. The shell progenitor masses in the Illustris simulation are on average bigger than masses of non shell-causing progenitors. The distribution peaks at around $\SI{1e11}{\Msun}$ \citep{pop+18}, which is slightly more massive than our median value. 

In conclusion, we find shell systems on average to be more massive than streams, which extends to their progenitors. Still, progenitors down to masses of $\log_{10}(M_\star/\Msun) \approx 9.6$ can form shells, making it more likely that the orbital configuration is the main parameter to predict shell-formation. The impact of the merger ratio will be examined in \cref{sec:ratio-gas-morphology}. We also provide an expectation for the mass recovery fraction, i.e. the mass fraction of a feature to its progenitor, to be 10-20\% for most features. 

\subsubsection{Half-light radius of streams vs. half-mass radius of progenitors }
\label{subsec:streampy}

\begin{figure}
    \centering
    \includegraphics[width = 0.45\textwidth]{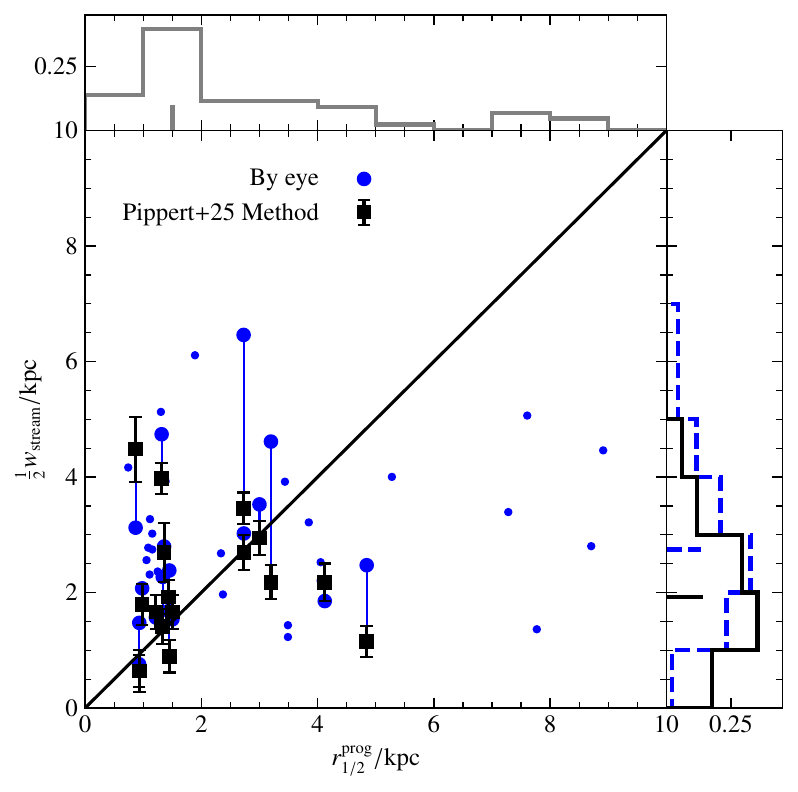}
    \caption{The width $w$ of streams that was determined by eye (blue circles) and a version determined by using \textsc{astrostreampy} (black squares) as a function of their progenitor's half-mass radius $r_{1/2}^\mathrm{prog}$. The black error bars represent the median and the standard deviation of the means within each box of the \textsc{astrostreampy} algorithm, which fits a Gaussian profile to the stream within multiple boxes at different positions along the stream. The \enquote{by eye} version is connected to its \textsc{astrostreampy} counterpart by a thin blue line. On the top, the distribution of the stellar half-mass radius, and to the right, the distributions of the widths determined by eye (blue dashed line) and of the widths determined via \textsc{astrostreampy} (black solid line) are shown. The short lines are the medians of each distribution.} 
    \label{fig: 8-sizes}
\end{figure}

\begin{figure*}
    \centering
    \includegraphics[width = \textwidth]{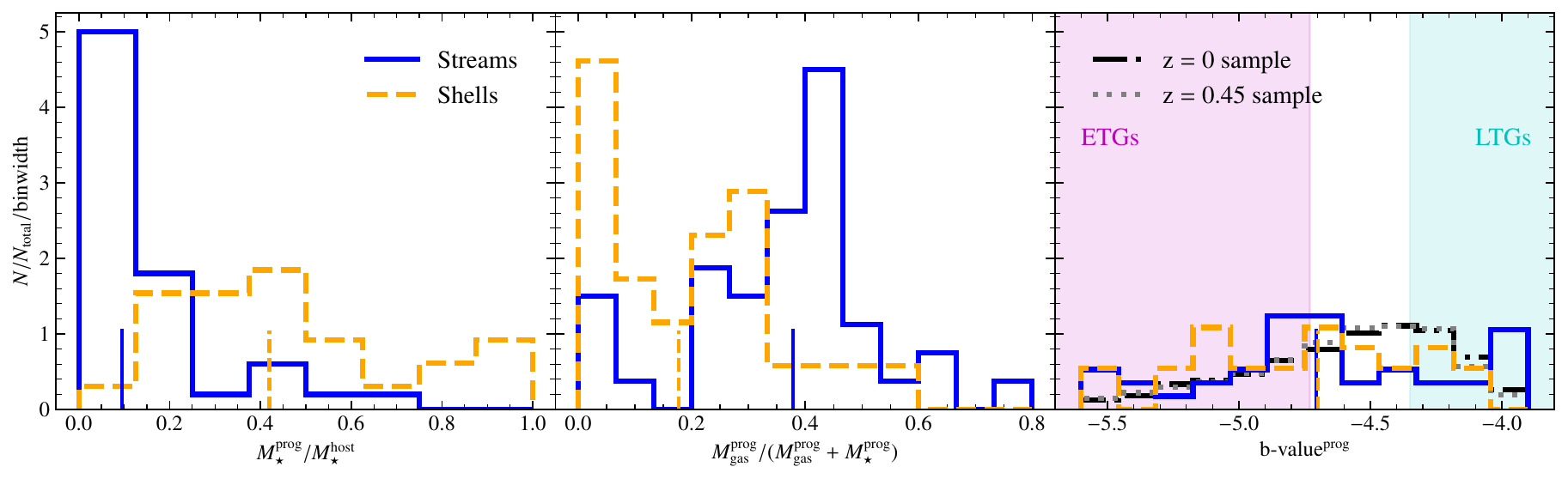} 
    \caption{ Left panel: The distribution of the stellar mass merger ratio $M_\star^\mathrm{prog}/M_\star^\mathrm{host}$ of the shell (orange dashed line) and stream (blue solid line) forming galaxy mergers. Middle panel: The distribution of the gas fraction $M_\mathrm{gas}^\mathrm{prog}/(M_\star^\mathrm{prog} + M_\mathrm{gas},^\mathrm{prog})$ of shell (orange dashed line) and stream (blue solid line) progenitors. Right panel: The distribution of the b-value$^\mathrm{prog}$ of shell (orange dashed line) and stream (blue solid line) progenitors, and the overall galaxy sample above $M_\star > \SI{1e9}{\Msun}$ at z = 0.07 (black dashed line) and z = 0.45 (gray dotted line), which is the median redshift of peak stellar mass for the progenitors. The vertical dashed colored lines show the respective median. The boundaries between spheroidal, intermediate, and disk galaxies are indicated by the magenta and cyan shaded areas. The masses of progenitors are calculated within a radius of $0.1\rvir$ at the redshift of peak stellar mass $z_\mathrm{PSM}$. The same is true for the masses of the host galaxies. The b-value is calculated within $3\rhalf$ at the redshift of peak stellar mass $z_\mathrm{PSM}$. }
    \label{fig: 9-prog_props}
\end{figure*}

Another fundamental property of galaxies is their size. After being disrupted and forming a stream, it is unclear whether the resulting stream will be broader or more compact than its progenitor. It might puff up due to induced random motion \citep{amorisco15} or being compacted due to being tidally elongated in the perpendicular direction, or both. \citet{errani+15} used isolated merger simulations to conclude that the half-light radius of a stream increases with distance from the host galaxy, which rather supports the first scenario. This agrees with the analytical model by \citet{erkal+16}. We calculate half-mass radii of the progenitors at their redshift of peak stellar mass and half-light radii of the streams extracted from a Gaussian fit to the stream using \textsc{astrostreampy}\footnote{\url{https://github.com/jnpippert/streampy},\url{https://pypi.org/project/astrostreampy}} by \citet{pippert+25}. Additionally, the widths $w_\mathrm{stream}$ as already presented in \cref{fig: 2-stats} are included. The half-light radii are based on the same mock observations that were used to identify the streams in the first place. To prepare the mock observations for the photometric stream modeling, we subtract a model of the host galaxy using the multi-Gaussian expansion code \textsc{MgeFit} by \citet{cappellari02}. Sources not belonging to the stellar halo of the host are masked beforehand. This enables a clearer vision of the feature and allows us to create corresponding masks for the stream modeling. The extracted stream and its mask are then passed to \textsc{astrostreampy} to model and retrieve shape properties such as the effective width $w_\mathrm{stream}$. For more details, we refer to \citet{pippert+25}. This process was successful for 17 streams. The median half-light radius of our sample using this method is $1.92_{-0.26}^{+0.62} \mathrm{\ kpc}$ (using the 0.32 and 0.67-quantile as scatter). We compare our findings to observations by \citet{pippert+25} who find a median half-light radius of $2.2_{-0.38}^{+0.70} \mathrm{\ kpc}$ (same scatter) for their sample, which is in good agreement, demonstrating that the method also works for mock observations of simulated streams. 

\cref{fig: 8-sizes} shows half the width of streams $w_\mathrm{stream}$ as calculated using the method by \citet{pippert+25} --\textsc{astrostreampy} -- (black dots with errorbars), and the visually inferred widths (blue dots) as a function of the progenitor's half-mass radius $r_{1/2}^\mathrm{prog}$. The black and blue dots representing the same progenitor are connected by a blue line. No clear correlation is visible. On average, the visually inferred half-widths are larger than the half-light radii calculated by \textsc{astrostreampy} (black dots), but especially at small progenitor radii, both methods are consistent with each other in absolute terms. There is a small cluster at small progenitor radii that is consistent with the one-to-one relation, suggesting that the width of the stream is roughly the same as the half-mass radius of the stream's progenitor. But the non-existence of a clear correlation is evidence that the exact position along the stream and its evolutionary stage at the point of measurement influence its width more strongly than the size of its progenitor, which is consistent with the findings by \citet{errani+15} and \citet{erkal+16}. This clearly shows that the progenitor sizes cannot be recovered from the observed streams' widths using photometry.  

\subsubsection{Merger Ratio, gas Fraction and morphology}
\label{sec:ratio-gas-morphology}

To further test the differences of the progenitors of shells and streams, we compare the stellar merger ratio $M_\star^\mathrm{prog}/M_\star^\mathrm{host}$, gas-mass fraction $M_\mathrm{gas}^\mathrm{prog}/(M_\mathrm{gas}^\mathrm{prog} + M_\star^\mathrm{prog})$ and a tracer for a galaxy's morphology between shell and stream progenitors. The morphology tracer used is the b-value \citep{teklu+15, teklu+17} given by

\begin{equation}
    b = \log_{10}\biggl( \frac{j_\star}{\text{kpc km/s}} \biggr) - \frac{2}{3} \log_{10}\biggl( \frac{M_\star}{M_\odot} \biggr)\,,
\end{equation}

where $j_\star$ is the stellar specific angular momentum and $M_\star$ the stellar mass. \citet{teklu+17} introduced the classification that galaxies with b\,$> -4.35$ are disks, while galaxies with b\,$< -4.73$ are spheroids, following the observational findings by \citet{romanowsky&fall12}. All galaxies in between are intermediates. 

Figure \ref{fig: 9-prog_props} shows the distribution of the stellar merger ratio (left panel), the distribution of the gas-mass fraction (middle panel), and the distribution of the b-value (right panel). It is evident that shell-forming mergers (orange) have high merger ratios ($0.2 < M_\star^\mathrm{prog}/M_\star^\mathrm{host} < 1$), which includes minor and major mergers, consistent with results found by \citet{pop+18} for shells (more than 50\% of shell-forming mergers in the Illustris simulation happen at $M_\star^\mathrm{prog}/M_\star^\mathrm{host} > 1/3$) and the results for shells by \citet{mancillas+19} from a cosmological simulation. Stream-forming mergers (blue) are mostly minor to mini mergers  ($M_\star^\mathrm{prog}/M_\star^\mathrm{host} < 0.2$), consistent with results for streams by \citet{mancillas+19}. 

The gas-mass fraction (central panel) of stream progenitors is on average higher than for shell progenitors. This is likely related to the stellar age of the stream, as a progenitor with a lot of gas is more likely to trigger star formation within the developing feature, causing it to be young. 

Finally, there is almost no difference in morphology (b-value) between shell and stream progenitors (right panel). Both cover the entire range from spheroidal to disk galaxies, but disky progenitors are more frequent for stream progenitors ($\sim 28\%$) than for shell progenitors ($\sim 19\%$). Both fractions are less than the fraction of disky galaxies in the overall galaxy sample ($\sim 31\%$) at z = 0 (black) and at $z = 0.45$ (gray), the median redshift of peak stellar mass for shell and stream progenitors. Therefore, a stream progenitor is slightly more likely to be a disk galaxy, than a shell progenitor is, which is likely related to the fact that spheroidal galaxies are on average more massive than disk galaxies, which is also the case for shell progenitors, as seen in \cref{fig: 7-masses}. The results by \citet{pop+18}, who found that shell progenitors tend to have a larger amount of rotational support than their overall population of accreted satellites, suggest that shell progenitors are slightly more disky (or rotation-dominated) than the overall satellite population. This seems contradicting to our results, as it seems unlikely that stream progenitors are even more disky than the total galaxy sample in Illustris. However, they do not compare this to a stream progenitor sample. 

In summary, we confirm the prior findings from isolated simulations by \citet{karademir+19} and \citet{amorisco15}, finding that, while major mergers are the dominant formation channel for shells, it is possible to make shells from minor mergers. As found by \citet{karademir+19}, the crucial component is the type of orbit, with a head-on collision necessary to make a shell. Consequently, progenitors of shells tend to be more massive than progenitors of streams, specifically because the frequency of head-on collisions is greater for the more massive mergers. This also explains why \citet{pop+18} find progenitors of shells to be major mergers. Additionally, there is a slight preference for stream progenitors to be disky compared to shell progenitors in our simulation.  

\subsubsection{Mass-Size and mass-metallicity relation}

\begin{figure}
    \centering
    \includegraphics[width = 0.4\textwidth]{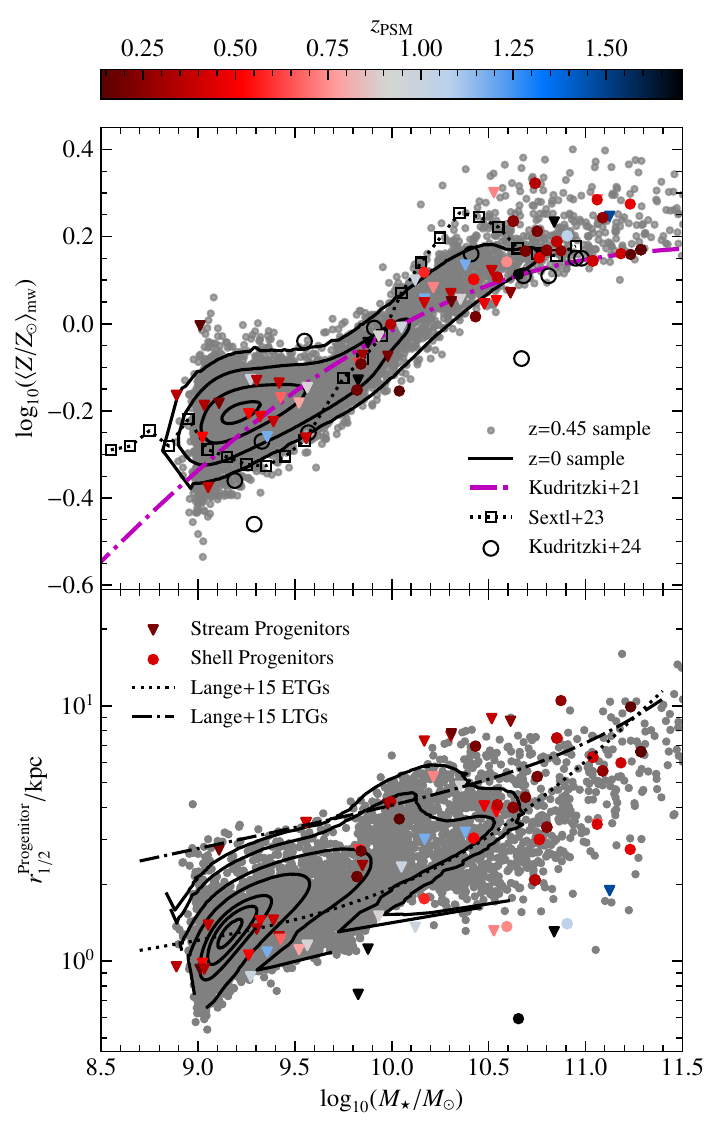} 
    \caption{Top panel: Mass-weighted metallicity $\log_{10}( \langle Z/Z_\odot \rangle_\mathrm{mw} )$ of shell (circle) and stream (triangle) progenitors as a function of stellar mass $M_\star^\mathrm{prog}$, colored by the redshift of peak stellar mass $z_\mathrm{PSM}$. Overplotted are results from stacked galaxy spectra by \citet{sextl+24} (open black squares and dotted line), from individual blue supergiants measured by \citet{kudritzki+24} (open black dots), as well as the prediction of a galaxy evolution lookback model by \citet{kudritzki+21} (magenta dash-dotted line). Bottom panel: The half-mass radius $\rhalf$ of shell (circle) and stream (triangle) progenitors as a function of stellar mass $M_\star^\mathrm{prog}$, colored by the redshift of peak stellar mass $z_\mathrm{PSM}$. Overplotted are observed fits (black lines) to the mass-size relation for ETGs (dotted) and LTGs (dashed) by \citet{lange+15}. The fitting parameters are taken from their population definition according to morphology in the r-band. In both panels, the gray dots represent the whole sample of galaxies at $z = 0.45$ and the black contours indicate the whole sample of galaxies at $z = 0$. All properties are calculated at the redshift of peak stellar mass $z_\mathrm{PSM}$.}
    \label{fig: 10-relations}
\end{figure}

In \cref{subsec:sigma-age-met} we present the stellar age and metallicity of shells and streams at $z = 0.07$. The top panel of \cref{fig: 10-relations} shows the mass-metallicity relation \citep[e.g.][]{gallazzi+05, zahid+17} for their progenitors, which can potentially explain the difference in the metallicity of the $z=0.07$ features. Shell progenitors (triangles) and stream progenitors (circles) are colored by their redshift of peak stellar mass, whose median is $ z_\mathrm{PSM} = 0.45$. Therefore, the whole galaxy sample from the Magneticum simulations at that redshift is shown as gray dots alongside the $z=0.07$ sample that is presented as black contours. All progenitors lie perfectly on top of the two comparison samples and are also consistent with observations through the use of individual blue supergiants \citep[][open black circles]{kudritzki+24}, a galaxy evolution lookback model \citep[][magenta dashed line]{kudritzki+21}, and measurements of the young stellar population using stellar population synthesis on stacked SDSS galaxies of star-forming galaxies \citep[][open black squares]{sextl+23}. This indicates that the progenitors of shells and streams do not occupy any special region in the mass-metallicity relation, but it is noteworthy that shell progenitors (triangles) primarily lie at the high-mass end and have high metallicities, while stream progenitors (circles) are at the low-mass end and have low metallicities. This suggests that the difference in metallicity between the current-day shells and streams arises simply from the fact that shell progenitors are more massive (as seen in \cref{fig: 7-masses}) and therefore are also more metal-rich than stream progenitors. Similarly, \citet{panithanpaisal+21} do not find a significant difference between stellar stream progenitors and present day simulated galaxies regarding the mass-metallicity relation within the FIRE simulations \citep{Hopkins+18}. However, the whole simulation underpredicts observed metallicities \citep{panithanpaisal+21}.

In addition, the bottom panel of \cref{fig: 10-relations} for shell and stream progenitors, presents the mass-size relation. The same comparison samples as in the bottom panel are shown, together with fits to the observed mass-size relation by \citet[][black dotted line for ETGs, black dash-dotted line for LTGs]{lange+15}. Again, the progenitors do not occupy a special region.

These two scaling relations show that the progenitors of shells and streams are not outliers among the overall galaxy population regarding their metallicity and size. For more details on scaling relations in the \emph{Magneticum Simulations}, we refer to \citet{dolag+25}. 

\section{In-situ streams}
\label{sec: in-situ streams}

\begin{figure*}
    \centering
    \includegraphics[width = \textwidth]{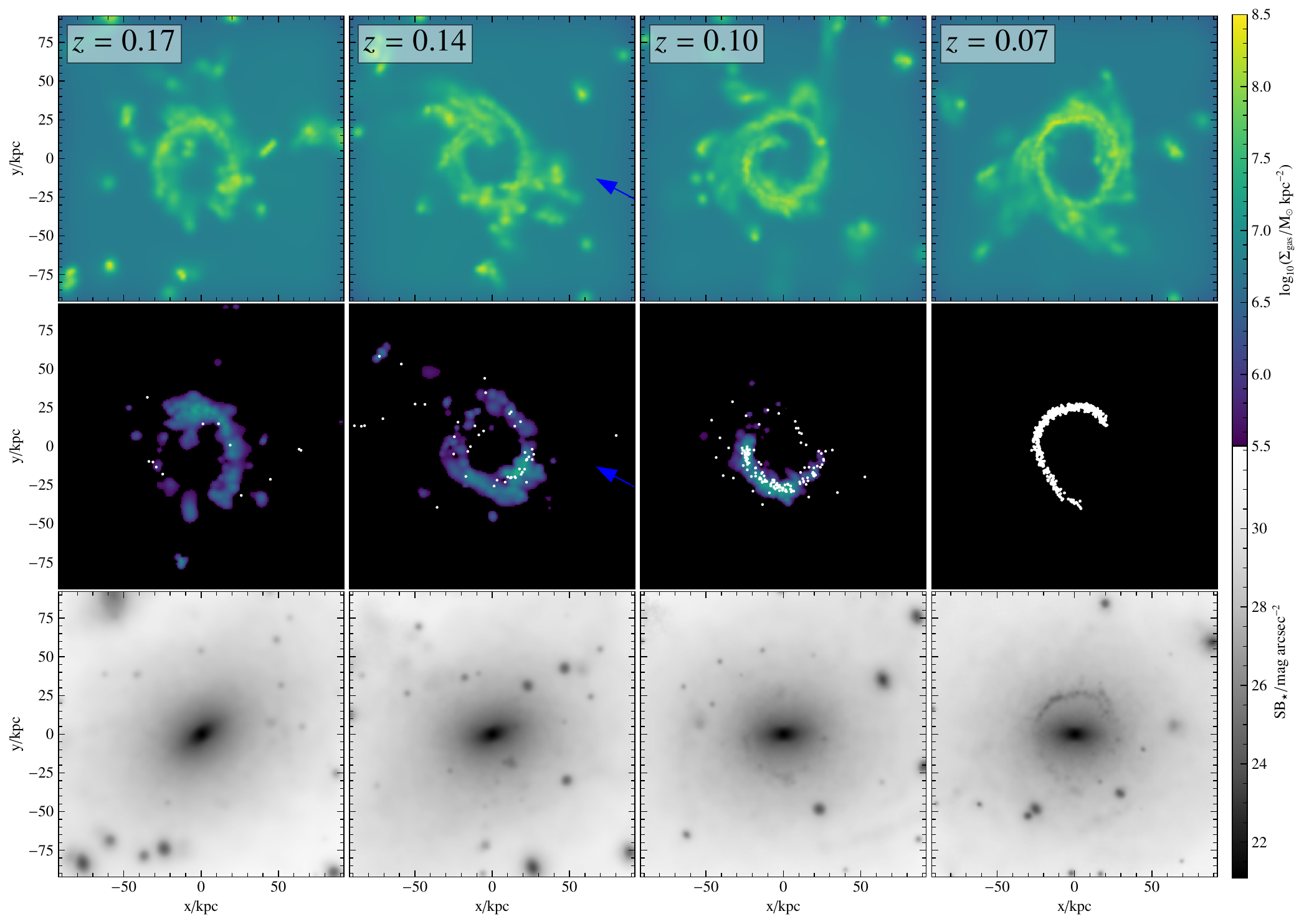} 
    \caption{Formation of an illustrative, in-situ stream not identified by the progenitor finding algorithm (UID 6070). The panels in the top row show the gas surface density map of the stream's host galaxy. The panels in the middle row show the stars (white dots) and the gas surface density map (colored) of the particles ending up being stars within the apparent stream at $z = 0.07$ and which are younger than $\SI{2.1}{\giga\year}$, which corresponds to the lookback time of $z = 0.17$. The host galaxy was removed in the middle row. The panels in the bottom row show the stellar mock maps of the host galaxy. Time evolves from the left ($z = 0.17$) to the right ($z = 0.07$). Each particle distribution is centered and rotated into the face-on frame of the host galaxy at $z = 0.07$. The blue arrow depicts the position and direction of the velocity of a satellite galaxy, which is a potential trigger for the star formation (see \cref{app:third_appendix}).}
    \label{fig: 11-young_stream}
\end{figure*}

Finally, we focus on the young ($\langle t_\star \rangle < \SI{7}{\giga\year}$) streams for which no progenitor could be identified. Directly tracing back the star particles selected within each stream should reveal where they came from. As the streams are young, it is likely that their star particles formed recently out of gas particles. For this reason, we matched the gas particles within each snapshot with the stellar particles of the stream. Note that in our star formation model, one gas particle forms up to four star particles \citep{dolag+25}. It is therefore possible that not only the direct gas particle progenitor of a star particle is selected, but also the particle representing the remaining gas after a star particle's formation. Figure \ref{fig: 11-young_stream} presents the evolution from $z = 0.17$ to $z = 0.07$ of the total gas content of the host galaxy (top row), the gas (green-yellow colored) and stars (white dots) that end up in or form the stars within an illustrative young stream (UID 6070, middle row), and the stellar mock observations of the host galaxy, from which the stream was identified (bottom right panel).

Within the host galaxy, a gas ring is visible over the whole evolution. The middle row shows a gas ring from $z = 0.17$ to $z = 0.14$, while some stars roughly align with this ring. At $z = 0.10$ the gas ring in the middle row becomes a gas semicircle, and newly formed stars appear within it. These stars were directly formed out of the gas ring present in the top row just shortly before this snapshot. Finally, all the stars are formed from this gas, which look like a stream in the last snapshot (at $z=0.07$) in the mock observation. Thus, the young arc-like feature that looks like a stream in our mock observation does not form from an infalling satellite galaxy, but rather directly forms out of the gas ring, which was established at high redshift. The details of the formation of this gas ring are beyond the scope of this work. It is, in other words, an in-situ stream. A similar formation was found for half of the eight streams younger than $\SI{7}{\giga\year}$ for which no progenitor was previously identified. A possible trigger for this star formation is a close encounter with a satellite galaxy that has a pericenter passage at $z = 0.14$. This potentially triggering satellite galaxy passes the host galaxy in the same direction as the star formation front (blue arrow, see \cref{app:third_appendix}). We propose this formation scenario for in-situ streams, which are features indistinguishable from a regular stream, except for their formation. A tell-tale sign is a very young age. Whether these are only found in simulations or are also present in current LSB surveys needs to be further investigated in the future. 

\section{Summary and conclusion}
\label{sec: conclusion}

We used the cosmological simulation Magneticum Box4 (uhr) to analyze and interpret the velocity dispersion, age, and metallicities of stellar shells and streams -- initially classified by \citet{valenzuela&remus24} -- as well as to identify their progenitor galaxies and to connect their properties to the present-day features. This is the first such complete analysis from a cosmological simulation combining shells and streams. We compared our results to shells around individual galaxies \citep{bilek+13, bilek+22}, annotated tidal features from deep CFHT images \citep{sola+22, sola+25}, and to the widths of streams found in the WWFI survey \citep{kluge+2020, pippert+25}.

Most shells and streams are identified in two out of three projections, in agreement with \citet{pop+18} (see \cref{tab: identify}) and predominantly in the face-on projection. We find that — in the face-on projection — the average projected radius of shells ($\sim \SI{20}{\kilo\parsec}$) is slightly smaller than for streams ($\sim \SI{27}{\kilo\parsec}$), but in agreement with observations by \citet{bilek+22} and \citet{sola+25}. The projected widths of shells ($\sim \SI{6.55}{\kilo\parsec}$) are slightly broader than for streams ($\sim \SI{5.55}{\kilo\parsec}$), which are in agreement with observations by \citet{pippert+25} and \citet{sola+25}. We found that most shells are aligned with the major axis because the shell-causing merger also elongates the host galaxy's shape, while the mean angular position of streams are homogeneously distributed. 

On average, shells and streams have a lower velocity dispersion than their surroundings, confirming a prediction by \citet{valenzuela&remus24}. This may be used to aid the identification of faint shells and streams in future observations. Ages and metallicities are on average similar to their surroundings. Noteworthy are a handful of very young streams.

By identifying their progenitor galaxies, we confirmed the notion of streams being formed by infalling satellite galaxies with a significant impact parameter, while shells are formed by a radial merger \citep[e.g.][]{quinn84,pop+18,karademir+19}. We found that shell progenitors are on average more massive than stream progenitors, which extends to their merger mass ratios, meaning that shells form predominantly from major mergers, while streams are formed by minor mergers \citep[e.g.][]{mancillas+19}. 

The progenitors of shells tend to be more massive than progenitors of streams, specifically because the frequency of head-on collisions is greater for the more massive mergers. Major mergers are not strictly necessary to form shells, because we also find shells that were formed by minor mergers, consistent with \citet{karademir+19} who identified the type of orbit as the crucial component of shell formation. 

We compared the widths of streams measured in mock observations to the half-mass radius of their progenitor galaxies and did not find a correlation, which suggests that progenitor sizes cannot be recovered from the observed streams’ widths using photometry. The evolutionary stage of a stream \citep[e.g.][]{errani+15} is more important for the local width than the progenitor's size.

The progenitors of shells and streams do not occupy a special region within the stellar mass-size relation and the stellar mass-metallicity relation, which grants further credibility to mass estimates of the progenitors of observed shells or streams based on this relation. Shells occupy the higher-mass end in the mass-metallicity relation, and therefore are also more metal-rich than stream progenitors. 
 
Finally, we further investigated prominent young stellar streams, for which no progenitor could be identified by cross-matching stellar particles. This kind of otherwise indistinguishable young stellar streams does not result from an infalling satellite galaxy, but rather forms in-situ out of the gas already existing within the galaxy. This star formation is likely triggered by a passing satellite galaxy shortly before the appearance of the stream. We propose this formation scenario for features that look like streams and are very young. Whether these are only found in simulations or are also present in current LSB surveys needs to be further explored. 

 In conclusion, our analysis of shells and streams within the Magneticum Box4 (uhr) simulation provides the first combined analysis from a cosmological simulation that links velocity dispersion, ages, metallicities, and structural properties to the properties of their progenitors. We further justify the practical diagnostics of the velocity dispersion contrasts to identify shells and streams in future LSB, but progenitor sizes appear difficult to infer from stream widths based on our current analysis. We provide an analysis of the formation of shells and streams, demonstrating that shell progenitors are on average more massive than progenitors of streams, with the merger orbits thus being more head-on. We further demonstrate a new type of young stream, formed in-situ from an existing host gas ring through an instability, potentially triggered by a fly-by. Further work, particularly on connecting young in-situ streams to observational data, will be essential to fully unravel the origins of tidal features and their role in shaping present-day galaxies.

\begin{acknowledgements}
We thank all the anonymous referee for useful comments that enhanced the quality of this manuscript. JS and KD acknowledge support by the COMPLEX project from the European Research Council (ERC) under the European Union’s Horizon 2020 research and innovation program grant agreement ERC-2019-AdG 882679. LMV acknowledges support by the German Academic Scholarship Foundation (Studienstiftung des deutschen Volkes) and the Marianne-Plehn-Program of the Elite Network of Bavaria. ES acknowledges the Leverhulme Trust for funding under the grant number RPG-2021-20. The calculations for the hydrodynamical simulations were carried out at the Leibniz Supercomputer center (LRZ) under the project pr83li (Magneticum). This research was supported by the Excellence Cluster ORIGINS, funded by the Deutsche Forschungsgemeinschaft under Germany's Excellence Strategy -- EXC-2094-390783311. The following software was used for this work: Julia \citep{bezanson+17:julia}, \textsc{matplotlib} \citep{hunter2007},  SPHtoGrid.jl and GadgetIO.jl \citep{boessvalenzuela24}, and \textsc{astrostreampy} \citep{pippert+25}. 
\end{acknowledgements}

\bibliographystyle{style/aa}
\bibliography{bib} 

\begin{appendix}

\onecolumn
\section{Diagnostic plots for the stellar population properties}
\label{app:first_appendix}

The visual selection of streams from the stellar mock observations by overlaying a polygon (cyan line) is illustrated in \cref{fig: a2-dia-stream}. This approach is similar to the annotation tool for LSB features \code{JAFAR}\footnote{\url{https://jafar.astro.unistra.fr/}} presented by \citet{sola+22}. Subsequently, the Voronoi bins lying within the polygon are used to calculate a median, 0.32- and 0.67-quantile as the population properties of the stream. The Voronoi bins within an annulus (white circles), whose inner and outer lines are the inner and outer radii of the stream, and which are not within the stream, are used to classify the surroundings of the stream. Additionally, the width of the stream $w_\mathrm{stream}$ is visually inferred by selecting two suitable vertices of the polygon. Finally, \cref{fig: a2-dia-stream} also shows a gas surface density map to get a broader view of the formation of the stream.

\begin{figure*}[htbp]
    \centering
    \includegraphics[width = \textwidth]{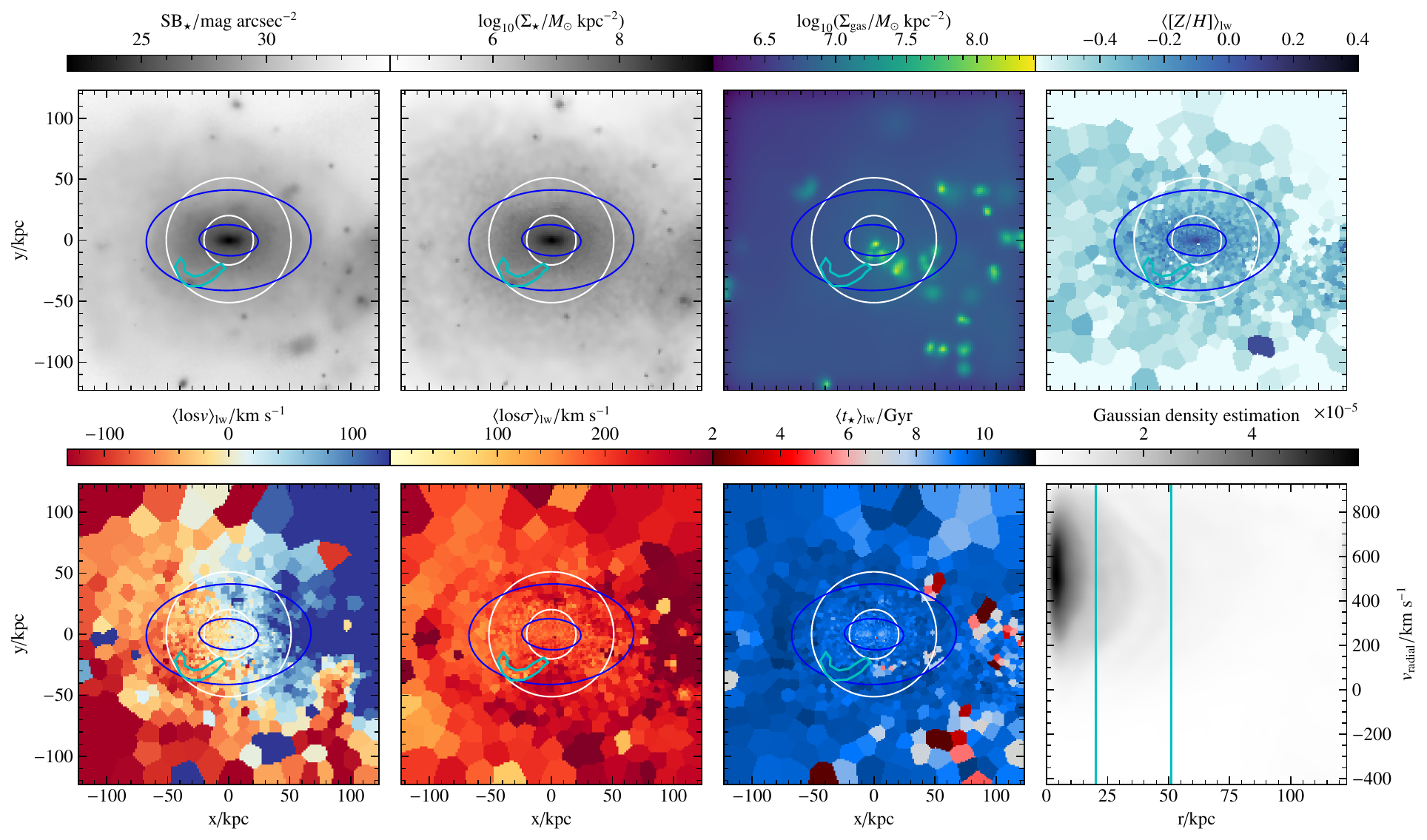}
    \caption{ Top row: Stellar mock surface brightness map, 2D-binned stellar mass map, 2D-binned gas mass map, and Luminosity-weighted Voronoi-binned maps of the stellar metallicity $\langle [Z/H] \rangle_\mathrm{lw}$. Bottom row: Luminosity-weighted Voronoi-binned maps of the line-of-sight velocity $\mathrm{los}v$, the velocity dispersion $\mathrm{los}\sigma$, and the stellar age $\langle t_\star \rangle_\mathrm{lw}$. All are in the face-on projection. The bottom right panel presents the radial velocity-radius phase space distribution of stellar particles colored by a Gaussian kernel density estimation. The cyan vertical lines are the inner and outer radius of the shell, while the blue vertical lines are $1\rhalf$ and $3\rhalf$. Overplotted on all maps is the polygon selecting the stream as cyan lines, the inner and outer circle of the  surrounding annulus in white, and the shape ellipses at $1\rhalf$ and $3\rhalf$ in blue.} 
    \label{fig: a2-dia-stream}
\end{figure*}

\onecolumn
\section{Overview of extracted stellar shells and streams from Magneticum}
\label{app:second_appendix}

\cref{fig: a3-mosaic_shells} presents our catalog of simulated shells and \cref{fig: a4-mosaic-streams} our catalog of simulated streams. They are produced in the same way as \cref{fig: 4-decomp}. The center of each panel is the center of the host galaxy (not shown), the side length of each panel is equal to nine times the half-mass radius of the host galaxy.

\begin{figure*}[htbp]
    \centering
    \includegraphics[width = 0.9\textwidth]{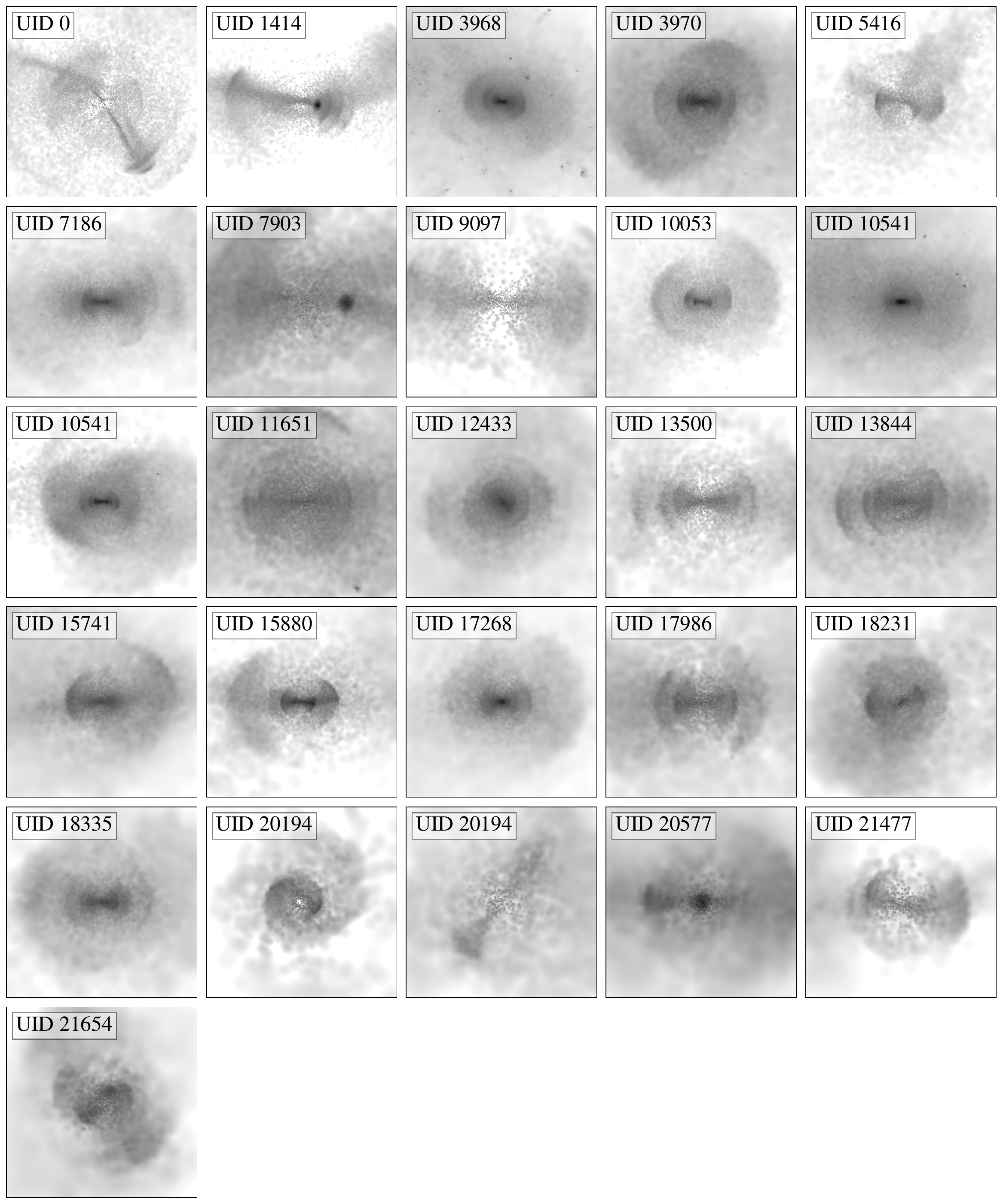}
    \caption{Stellar surface density maps of the particles which used to belong to a common subhalo, that was identified to be the progenitor of the shells, traced forward to $z=0.07$ for all shell progenitors.}
    \label{fig: a3-mosaic_shells}
\end{figure*}

\begin{figure*}[htbp]
    \centering
    \includegraphics[width = \textwidth]{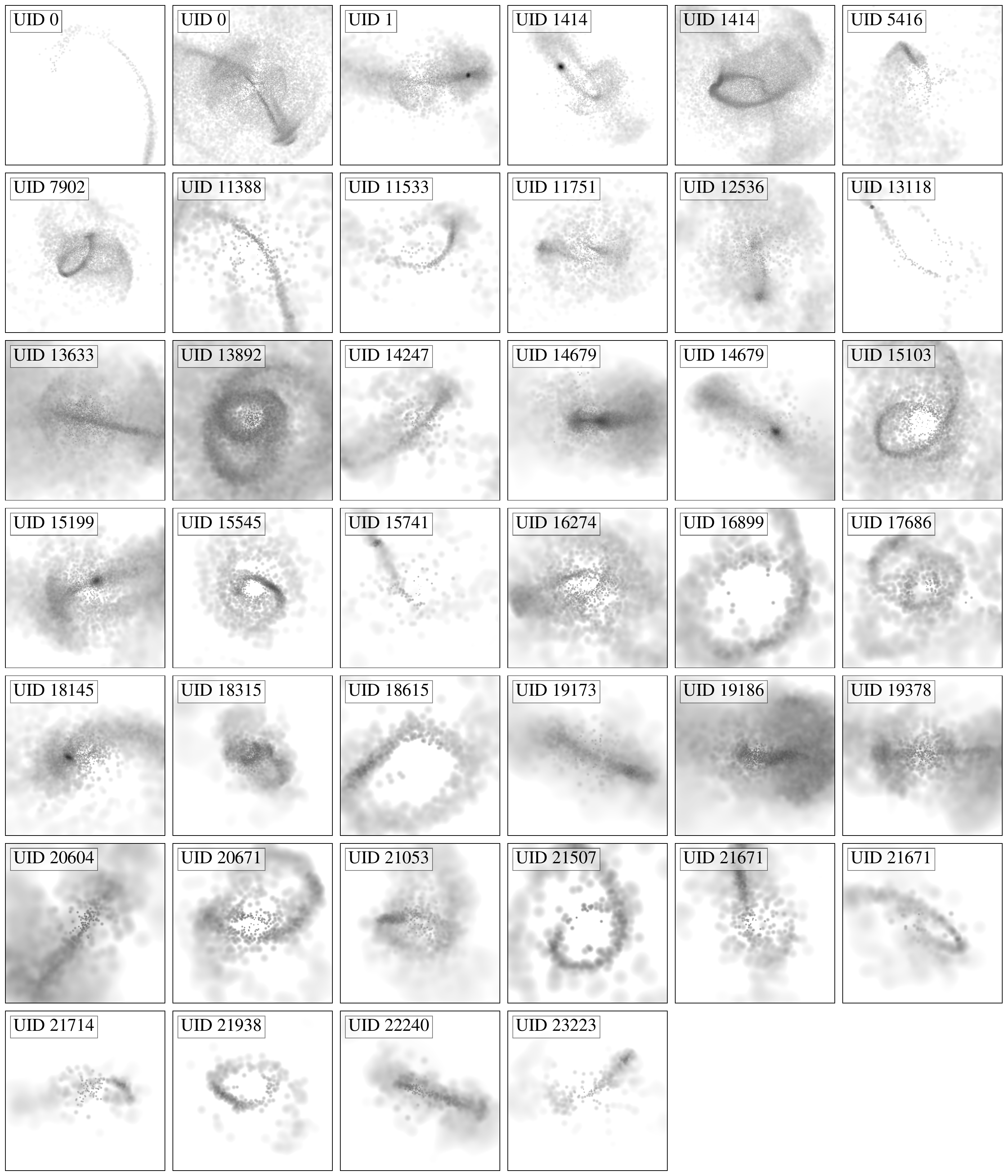}
    \caption{ Stellar surface density maps of the particles which used to belong to a common subhalo, that was identified to be the progenitor of the stream, traced forward to $z=0.07$ for all stream progenitors.}
    \label{fig: a4-mosaic-streams}
\end{figure*}

\onecolumn
\section{Stellar mock observations of host galaxies}
\label{app:last_appendix}

\begin{figure*}[htbp]
    \centering
    \includegraphics[width = 0.9\textwidth]{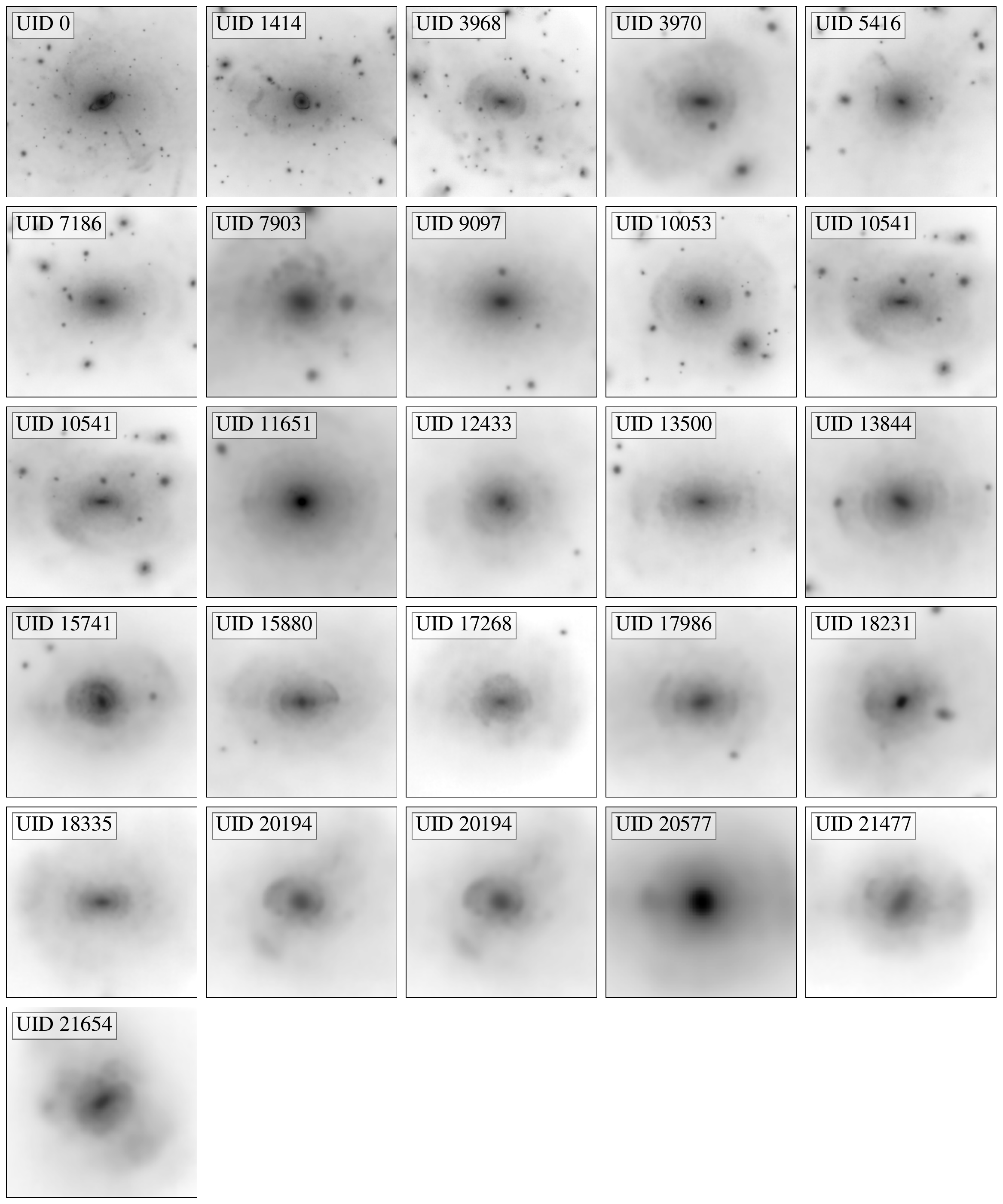}
    \caption{Stellar mock observations of each host galaxy for which a shell progenitor was identified. As this is a mirror of \cref{fig: a3-mosaic_shells}, some host galaxies are doubled so that it is easier to identify the corresponding panel in \cref{fig: a3-mosaic_shells}.}
    \label{fig: a3-mosaic_shells_host}
\end{figure*}

\begin{figure*}[htbp]
    \centering
    \includegraphics[width = \textwidth]{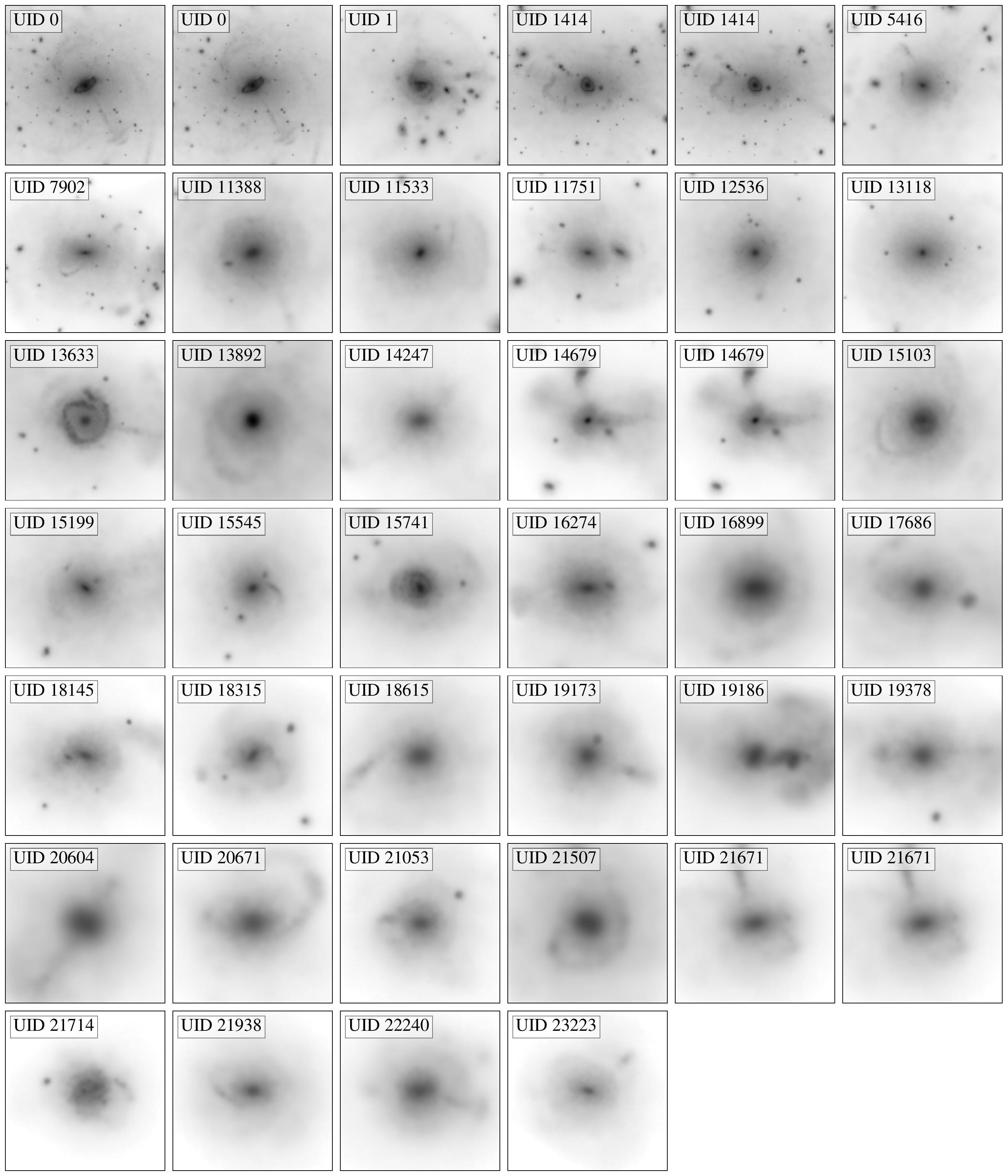}
    \caption{Stellar mock observations of each host galaxy for which a stream progenitor was identified. As this is a mirror of \cref{fig: a4-mosaic-streams}, some host galaxies are doubled so that it is easier to identify the corresponding panel in \cref{fig: a4-mosaic-streams}.}
    \label{fig: a4-mosaic-streams_host}
\end{figure*}

\newpage 
\clearpage
\newpage 
\section{Close encounter potentially triggering the formation of an in-situ stream}
\label{app:third_appendix}

Section\,\ref{sec: in-situ streams} presents the formation scenario for features that look like streams for which no progenitor satellite could be identified by directly matching stellar particles (see \cref{sec:ident-progs}). Additionally, these features are young ($\langle t_\star \rangle < \SI{7}{\giga\year}$) compared to the remaining sample (see \cref{subsec:sigma-age-met}). Figure\,\ref{fig: 11-young_stream} shows that these streams form directly out of the gas present in the host galaxy. This star formation is potentially triggered by a close encounter with a satellite galaxy, which is presented in \cref{fig: a5-trigger}. The evolution of the stellar surface density is shown from $z = 0.17$ to $z = 0.07$ (from the left to the right), similar to \cref{fig: 11-young_stream}, in the face-on projection (top row) and the edge-on projection (bottom row). The first pericenter passage is roughly at $z = 0.14$, which is roughly $0.5$ Gyr before the in-situ stream starts to form at $z=0.1$ (see \cref{fig: 11-young_stream}). Therefore, we suspect this close encounter triggered the formation of the illustrative in-situ stream. 

\begin{figure*}[htbp]
    \centering
    \includegraphics[width = \textwidth]{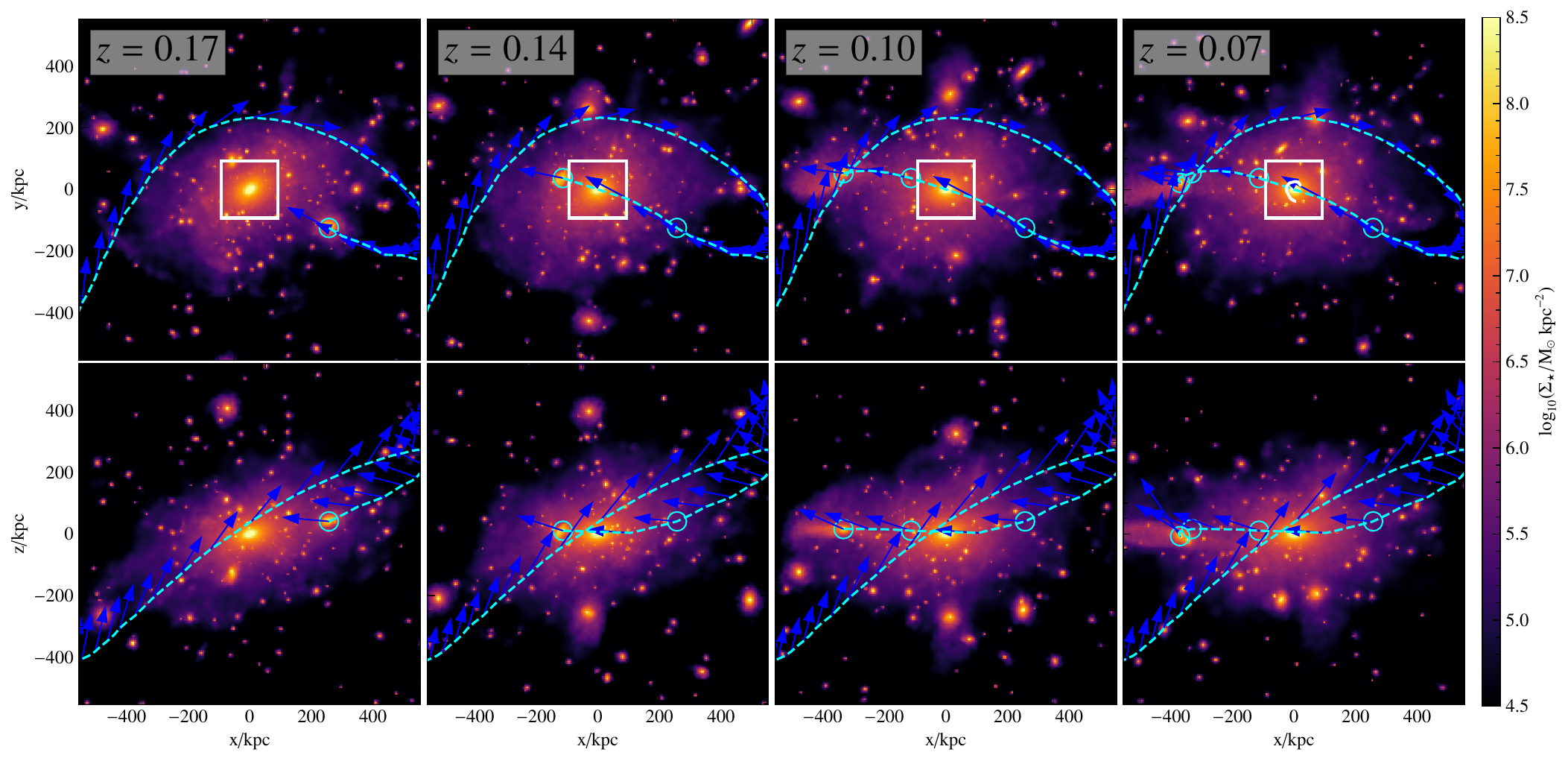}
    \caption{ The orbit of a satellite galaxy around the host galaxy of the in-situ stream, whose close encounter possibly triggered the in-situ stream's formation. The panels show the stellar surface density map of the stream's host galaxy face-on (top row) and edge-on (bottom row). The cyan dashed line depicts the orbit of the satellite galaxy and the blue arrows depict the direction of the velocity vector as identified by \textsc{Subfind} and \textsc{L-BaseTree}. The cyan circles enclose the satellite galaxy at each shown snapshot. The white frame in the top row shows the extent of \cref{fig: 11-young_stream} and the in-situ stream is indicated by the white contours within the white box of the top right panel.}
    \label{fig: a5-trigger}
\end{figure*}

\end{appendix}

\end{document}